\documentclass[twocolumn,times]{aastex61}

\newcommand{\arttwo}{\mbox{\sc art}$^2$}
\newcommand{\sunrise}{\mbox{\sc sunrise}}
\newcommand{\skirt}{\mbox{\sc skirt}}
\newcommand{\radishe}{\mbox{\sc radishe}}
\newcommand{\dartray}{\mbox{\sc dartray}}
\newcommand{\dirty}{\mbox{\sc dirty}}
\newcommand{\grasil}{\mbox{\sc grasil}}

\newcommand{\hyperion}{\mbox{\sc hyperion}}
\newcommand{\pd}{\mbox{\sc powderday}}

\newcommand{\agora}{\mbox{\sc agora}}

\newcommand{\gizmo}{\mbox{\sc gizmo}}
\newcommand{\gasoline}{\mbox{\sc gasoline}}
\newcommand{\changa}{\mbox{\sc changa}}

\newcommand{\arepo}{\mbox{\sc arepo}}
\newcommand{\enzo}{\mbox{\sc enzo}}

\newcommand{\fsps}{\mbox{\sc fsps}}

\newcommand{\yt}{\mbox{\sc yt}}

\received{\today}
\revised{}
\accepted{}
\submitjournal{ApJ}

\usepackage{color,comment}
\usepackage[normalem]{ulem}

\newcommand{\angstrom}{\text{\normalfont\AA}}

\shorttitle{}
\shortauthors{Narayanan et al.}

\begin{document}

\title[\pd: Code design, algorithms, and use cases]{\pd: Dust Radiative Transfer for Galaxy Simulations}

\correspondingauthor{Desika Narayanan}
\email{desika.narayanan@ufl.edu}

\author[0000-0002-7064-4309]{Desika Narayanan}
\affil{Department of Astronomy, University of Florida, 211 Bryant Space Sciences Center, Gainesville, FL 32611 USA}
\affil{University of Florida Informatics Institute, 432 Newell Drive, CISE Bldg E251, Gainesville, FL 32611}
\affil{Cosmic Dawn Center at the Niels Bohr Institute, University of Copenhagen and DTU-Space, Technical University of Denmark}
\author{Matthew J. Turk}
\affil{School of Information Sciences, University of Illinois, Urbana-Champaign, IL, 61820, USA}
\affil{Department of Astronomy, University of Illinois, Urbana-Champagin, IL, 61820, USA}
\author[0000-0002-8642-1329]{Thomas Robitaille}
\affil{Aperio Software, Headingley Enterprise \& Arts Centre, Bennett Road, Leeds LS6 3HN, UK}
\author[0000-0003-3850-4469]{Ashley J. Kelly}
\affil{Institute for Computational Cosmology, Department of Physics, Durham University, South Road, Durham, DH1 3LE UK}
\author[0000-0002-6040-8281]{B. Connor McClellan}
\affil{Department of Astronomy, University of Virginia, 530 McCormick Road, University of Virginia, Charlottesville, VA, 22904, USA}
\author[0000-0001-8350-4535]{Ray S Sharma}
\affil{Rutgers, the State University of New Jersey, 136 Frelinghuysen Road, Piscataway, NJ, 08854}
\author{Prerak Garg}
\affil{Department of Astronomy, University of Florida, 211 Bryant Space Sciences Center, Gainesville, FL 32611 USA}
\author{Matthew Abruzzo}
\affil{Department of Astronomy, Columbia University, New York, NY, 10027}
\author[0000-0002-8131-6378]{Ena Choi}
\affil{Quantum Universe Center, Korea Institute for Advanced Study, Hoegiro 85, Seoul 02455, Korea}
\author[0000-0002-1590-8551]{Charlie Conroy}
\affil{Center for Astrophysics $|$ Harvard \& Smithsonian, 60 Garden Street, Cambridge, MA 02138, USA} 
\author[0000-0002-9280-7594]{Benjamin D. Johnson}
\affil{Center for Astrophysics $|$ Harvard \& Smithsonian, 60 Garden Street, Cambridge, MA 02138, USA}
\author{Benjamin Kimock}
\affil{Department of Astronomy, University of Florida, 211 Bryant Space Sciences Center, Gainesville, FL 32611 USA}
\author{Qi Li}
\affil{Department of Astronomy, University of Florida, 211 Bryant Space Sciences Center, Gainesville, FL 32611 USA}
\author[0000-0001-7964-5933]{Christopher C. Lovell}
\affil{Centre for Astrophysical Research, School of Physics, Astronomy and Mathematics, University of Hertfordshire, Hatfield, AL10 9AB, UK}
\author[0000-0003-4422-8595]{Sidney Lower}
\affil{Department of Astronomy, University of Florida, 211 Bryant Space Sciences Center, Gainesville, FL 32611 USA}
\author[0000-0003-3474-1125]{George C. Privon}
\affil{Department of Astronomy, University of Florida, 211 Bryant Space Sciences Center, Gainesville, FL 32611 USA}
\affil{National Radio Astronomy Observatory, 520 Edgemont Road, Charlottesville, Va, 22903}
\author{Jonathan Roberts}
\affil{Department of Astronomy, University of Florida, 211 Bryant Space Sciences Center, Gainesville, FL 32611 USA}
\author{Snigdaa Sethuram}
\affil{Center for Relativistic Astrophysics, School of Physics, Georgia Institute of Technology, 837 State Street, Atlanta, GA 30332 USA}
\affil{Rutgers, the State University of New Jersey, 136 Frelinghuysen Road, Piscataway, NJ, 08854}
\author{Gregory F. Snyder}
\affil{Space Telescope Science Institute, 3700 San Martin Dr., Baltimore, MD, 21218}
\author{Robert Thompson}
\affil{Portalarium, 3410 Far West Blvd, Austin, TX, 78731}
\author[0000-0003-1173-8847]{John H. Wise}
\affil{Center for Relativistic Astrophysics, School of Physics, Georgia Institute of Technology, 837 State Street, Atlanta, GA 30332 USA}




\begin{abstract}
We present {\sc powderday}\footnote{Available at
  \url{https://github.com/dnarayanan/powderday}}, a flexible, fast,
open-source dust radiative transfer package designed to interface with
both idealized and cosmological galaxy formation simulations.  {\sc
  powderday} builds on {\sc fsps} stellar population synthesis models,
{\sc hyperion} dust radiative transfer, and employs {\sc yt} to
interface between different software packages.  We include our stellar
population synthesis modeling on the fly, which allows for significant
run-time flexibility in the assumed stellar physics, including the
initial mass function, stellar isochrone and spectra models, as well as in the assumed physics describing post-main sequence evolution. We include
a model for nebular line emission that can employ either pre-computed
{\sc cloudy} lookup tables (for efficiency), or direct photoionization
calculations for all young stars (for flexibility in \ion{H}{2} region
physics).  The dust content follows either simple
observationally-motivated prescriptions (i.e. constant dust to metals
ratios, or dust to gas ratios that vary with metallicity), direct
modeling from galaxy formation simulations that include  dust
physics, as well as a novel approach that includes the dust content
via learning-based algorithms from the {\sc simba} cosmological galaxy
formation simulation.  Active galactic nuclei (AGN) can additionally
be included via a range of prescriptions.  The output of these models
are broadband (912\AA{} -- 1mm) spectral energy distributions (SEDs), as
well as filter-convolved monochromatic images.  {\sc powderday} is
designed to eliminate last-mile efforts by researchers that employ
different hydrodynamic galaxy formation models, and seamlessly interfaces with {\sc
  gizmo}, {\sc arepo}, {\sc gasoline}, {\sc changa}, and {\sc enzo}.
We demonstrate the capabilities of the code via three applications: a
model for the star formation rate (SFR) - infrared luminosity relation
in galaxies (including the impact of AGN); the impact of circumstellar
dust around AGB stars on the mid-infrared
emission from galaxy SEDs; and the impact of galaxy inclination angle
on dust attenuation laws.

\end{abstract}




\section{Introduction}
\label{section:introduction}

The turn of the century ushered in dramatic advances in our knowledge
of cosmological galaxy evolution.  The advent
of medium and ultra-deep surveys across the electromagnetic spectrum
have resulted in the discovery of tens of thousands of galaxies
through the first billion years after the Big Bang
\citep[e.g.][]{shapley11a,madau14a,finkelstein16a,stark16a}.
These include populations of star-forming and passive galaxies at $z
\sim 2$ identified via novel color selection techniques
\citep{steidel96a,daddi04a,vandokkum08b}, galaxies at redshifts as
large as $z \sim 10$ \citep{finkelstein13a,finkelstein15a,oesch15a,oesch18},
and large samples of dusty starburst galaxies selected in the infrared
and submillimeter \citep*{blain02a,casey14a,lutz14a,hodge20a}.  Similarly,
targeted surveys of nearby galaxies have increased our understanding
both of their resolved stellar populations, as well as their molecular
and dusty interstellar medium properties
\citep[e.g.][]{kennicutt03b,kennicutt11a,dalcanton12a}.  These surveys
near and far have placed strong constraints on the cosmic evolution of
the star formation rate density, stellar mass, and the interstellar
medium (ISM) content in galaxies
\citep{blanton09a,kennicutt12a,madau14a,carilli13a}.


At the same time, simulations of galaxy formation have
become increasingly sophisticated, and shown substantial
progress in their ability to reproduce and interpret observations
\cite[see the recent reviews by][]{benson10a,somerville14a,naab17a,vogelsberger20a}.  These
simulations suggest a variety of mechanisms for shaping the physical
properties of galaxies at different mass scales, including black hole
growth and feedback, radiative feedback, gas accretion from the
intergalactic medium, and supernova-driven feedback amongst many
others.  Promisingly, despite the diverse range of methods used,
cosmological galaxy formation simulations have converged on a number
of predicted physical properties, including their predicted stellar
mass functions, SFR-$M_*$ relations, dust mass functions, and global gas fractions
\citep{dave12a,dave13a,dave19a,schaye14a,vogelsberger14a,somerville15a,hopkins18a,li19a}.  Of
course, in detail the physical properties of modeled galaxies are
strongly dependent on prescriptions for physical processes on small
scales such as star formation, black hole growth (and their associated
feedback), chemistry, the structure of the ISM, and so on.  Seemingly
small differences in any given prescription can have dramatic effects
on the observed properties of galaxies \citep[e.g.][]{hopkins13e}.

In order to quantitatively compare between numerical simulations of
galaxy formation and observations, one either needs to convert
integrated observational measures into physical quantities output by
the simulations, or map the physical properties generated in
simulations to bona fide observables.  The former method typically
relies on some sort of theoretical or empirical underpinning relating
observed quantities to physical properties, which can introduce some
level of uncertainty.  For example, uncertainties in galaxy star
formation histories, stellar evolution tracks, obscuring dust
geometries and the initial mass function propagate to uncertainties in
derived star formation rates and stellar masses of observed galaxies
\citep[e.g.][]{maraston06a,walcher11a,michalowski09a,conroy09b,conroy13a,dunlop11a,leja17a,zhang17a,leja19a,lower20a}.
Similarly, uncertainties in SED modeling, or the conversion between
 emission line strengths, continuum strengths, and gas masses are present in any
measurement of the ISM properties of galaxies
\citep[e.g.][]{casey12a,feldmann11a,narayanan11b,narayanan12a,bolatto13a,scoville14a,privon18a,liang18a}.

As reviewed by \citet*{steinacker13a} one alternative to
this is to utilize dust radiative transfer simulations to directly
calculate observables from the physical properties of galaxy formation
models.  To do this requires generating spectral energy
distributions for luminous sources, and modeling the transfer of this
radiation through the the interstellar medium.  The application of
dust radiative transfer models to galaxy simulations has a rich
history.  Indeed, a handful of both proprietary and open source codes
exist in the literature, including \sunrise
\ \citep{jonsson06a,jonsson10a,jonsson10b}, \skirt \ \citep{baes11a},
\radishe \ \citep{chakrabarti09a}, \, \dartray \ \citep{natale14a},
\dirty \ \citep{gordon01a,misselt01a}, \grasil \ \citep{dominguez13a}, {\sc radmc3d} \citep{dullemond12a}, 
and \arttwo \ \citep{li20a}.  \hyperion \ \citep{robitaille11a} is a
flexible and generic dust radiative transfer code that, while not
written specifically for galaxy formation simulations, can be used for
them.  We will discuss this particular code in much more detail later
in this paper.

Beyond serving as a tool for assessing how realistic modeled galaxies
are \citep[via, comparing their modeled morphologies or broadband
  colors to
  observations;][]{snyder15a,snyder15b,snyder17a,rodriguezgomez19a,abruzzo18a,schaye14a,torrey15a,narayanan11a,law12a,blecha18a}
computational galaxy formation studies that have employed dust
radiative transfer models have typically been used for two
purposes. The first is to understand the physical properties and
formation mechanisms of particular galaxy populations \citep[e.g.][]{efstathiou00a,granato00a,granato15a,younger09a,baugh05a,cen14a,chakrabarti07a,chakrabarti08a,cowley15a,narayanan09a,narayanan10a,narayanan10b,hayward11a,hayward13a,narayanan15a,snyder11a,blecha18a,kulier19a,mcalpine19a}, as well as galaxy model verification \citep[e.g.][]{trayford17a,camps18a,cochrane19a,baes19a,baes20a}.


A second powerful way to utilize dust radiative transfer models is to
investigate the ability of an observational tool in deriving physical
quantities.  For example, recent studies have investigated
quantitative morphology measures
\citep{lotz10a,lotz10b,snyder14a,abruzzo18a,snyder19a,cochrane19a}, star formation
rate indicators \citep{delooze14a,hayward14a}, stellar masses
\citep[][]{torrey15a,baes19a,katsianis20a,lower20a}, 
active galactic nuclei diagnostics \citep{snyder13a,narayanan10b}, bulge-disk
decomposition \citep{scannapieco10a}, the stellar initial mass function \citep[e.g.][]{baugh05a,narayanan12b,narayanan13b,cowley19a}, dust temperatures \citep{liang18a,liang19a,privon18a,ma19a} and galaxy dust attenuation curves \citep{narayanan18a,narayanan18b,ma19a,trayford20a}.


Despite the fact that dust radiative transfer codes have existed in
the literature for more than a decade, their usage with galaxy
formation simulations is only becoming common-place in the last few
years.  This is due, in part, to the fact that they can be
computationally demanding to run, as well as contain significant
`last-mile' efforts often needed to ensure correct data formats,
units, and model parameters that can vary significantly with different
hydrodynamic codes.  There is a need, in our view, for a dust
radiative transfer package with several attributes that can overcome
this.  First, such a code would optimally be highly flexible and
modular in its ability to vary critical quantities that impact the
simulated spectral energy distribution.  This might include the
stellar initial mass function, AGN emission model, and properties of the
stellar population synthesis model (such as the inclusion of thermally
pulsating asymptotic giant branch stars).  Second, a high level of
scalability is important.  Given ever-increasing mass and spatial
resolution in galaxy formation simulations
\citep[e.g.][]{hopkins17b,schaye14a,vogelsberger14a,dave19a}, the
ability to run efficient parallelized models is important.  Third,
there is a need for a portable code that interfaces with disparate
galaxy formation models.  Many of the existing codes in the literature
are optimized for a single hydrodynamic code, making it difficult to
compare between different data sets.  Comprehensive efforts to compare
cosmological and idealized hydrodynamic galaxy formation simulations
such as the \agora \ comparison project \citep{kim14a,kim16a} further
underscore the need for such a code package.

Motivated by this, as well as our vision for an open-source, community
supported dust radiative transfer package for galaxies, we build off
of previous efforts in this work and introduce \pd.  Our principle
goals with this code are to develop a lightweight, highly flexible and 
modular dust radiative transfer
package that interfaces seamlessly with numerous open-source galaxy
formation codes.  To achieve this, we build \pd \ on extremely
flexible packages, including \fsps \ for stellar population synthesis
\citep{conroy09b,conroy10a,conroy10b}, \hyperion \ for the Monte Carlo
radiative transfer \citep{robitaille11a}, and \yt \ for interfacing
with galaxy models \citep{turk11a}.

In this paper, we present the first release of \pd.  We outline the
basic algorithms and code methodologies, describe its usage, and
present examples highlighting the utility of a flexible dust radiative
transfer package.  The current code is currently designed to interface
seamlessly with outputs from  \gizmo
\ \citep[][]{hopkins14d,hopkins17a}, \gasoline \ \citep{wadsley04a}, \changa
\ \citep{menon15a}, \arepo \ \citep{springel10a} and \enzo \ \citep{bryan14a,brummel19}.
Finally, we close with an outlook to future directions
for development.  \pd \ itself is written in Python, though makes use
of Fortran, C and Cython via dependency software.

\section{Code Description -- Physics}
\label{section:code_description_physics}

\begin{figure*}
  \includegraphics[scale=0.5]{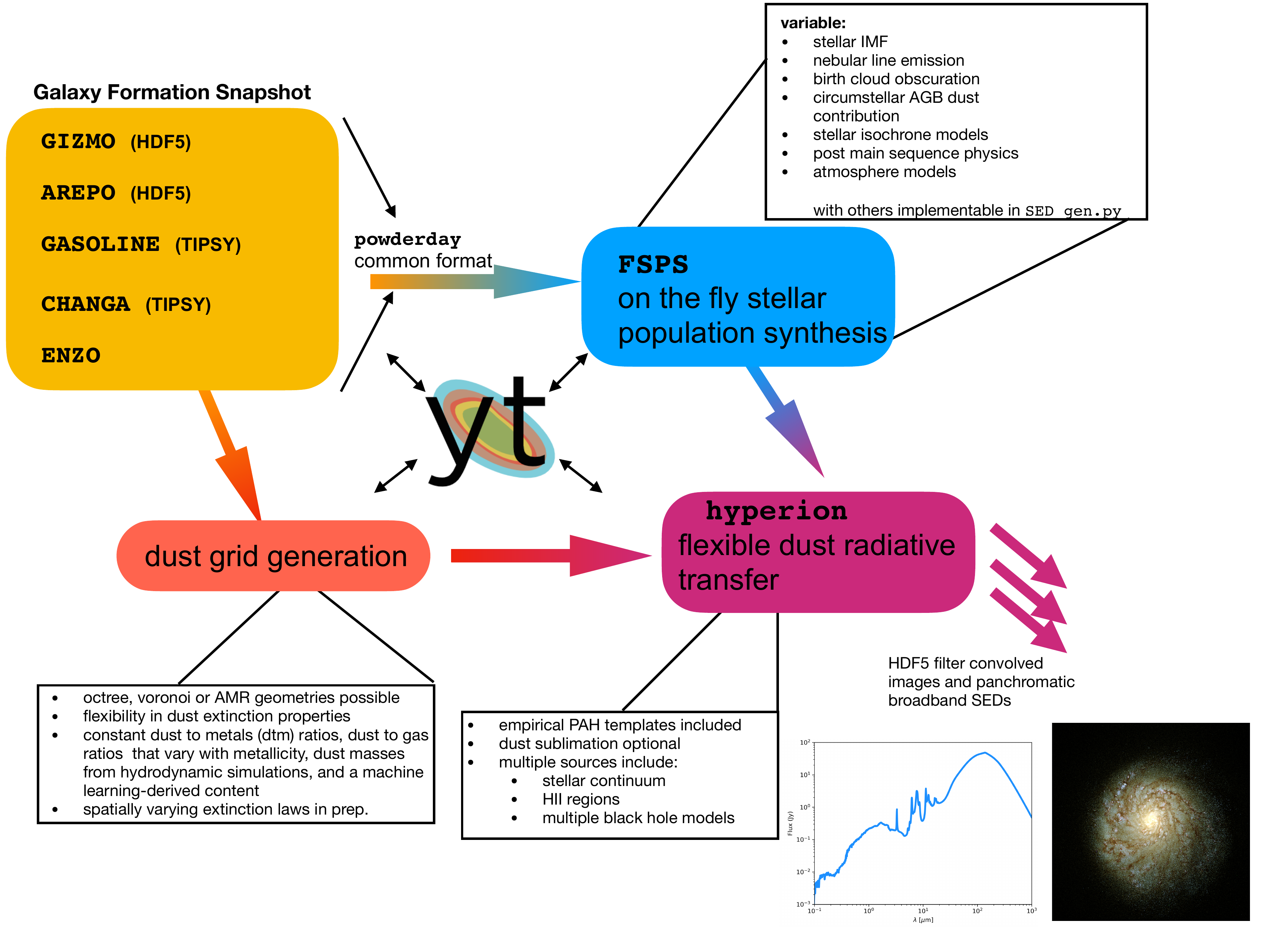}
  \caption{\label{figure:overview_schematic}{Schematic showing a high
      level view of the code architecture of {\sc powderday}.  \pd
      \ interfaces with a broad range of hydrodynamic galaxy evolution
      codes, regularizing them into a common format.  The stellar
      population synthesis is done on the fly with \fsps, and thus
      offers significant run-time flexibility.  Similar levels of
      flexibility exist with the dust grid generation, including the
      ability to use the outputs from newer on-the-fly dust evolution
      models in cosmological simulations.  Finally, the dust radiative
      transfer is performed with the \hyperion \ Monte Carlo dust
      radiative transfer code.  Throughout the entire code flow, \pd
      \ depends on \yt \ as a glue connecting a wide range of modules.
  }}
\end{figure*}

We begin the description of the code by first outlining the underlying
physics that goes into the radiative transfer.  We follow this in
\S~\ref{section:implementation} with a description of the
implementation itself. In Figure~\ref{figure:overview_schematic}, we
show a schematic of the overall code flow that will serve as a
reference throughout both this section, as well as in
\S~\ref{section:implementation}.

\subsection{Overview}

As a higher level overview: \pd \ projects the physical quantities
from hydrodynamic galaxy formation simulations onto an adaptive grid (or uses the underlying mesh, if available),
calculates the spectral energy distribution (SED) for the luminous
sources, and then propagates this light through the dusty interstellar
medium.  The dust temperatures are calculated self-consistently, so
that the final result from this are model SEDs from the ultraviolet
(longward of $912 \angstrom$) through millimeter wave.  In what
follows, we outline the details of these calculations.  This section
is not meant to serve as a user manual, but rather an overview of the
design and methods.  A full user manual can be found at
\url{https://powderday.readthedocs.io/en/latest/}.

\begin{table*}
        \centering
        \caption{Model Simulations Used in this Paper }
        \label{table:models}
        \begin{tabular}{lccccc}
                \hline
                Name & Type of Simulation & Type of Galaxy & Snapshot Type & Snapshot Location & Citation\\
                \hline
                     {\sc gizmodisk} & Cosmological Zoom & $z=0$ Disk & {\sc gizmo} HDF5 & \footnote{\url{https://users.astro.ufl.edu/~desika.narayanan/powderday_files/mufasa_gizmo_snapshot_134.hdf5}} &\citet{narayanan18a,narayanan18b} \\
                    {\sc latte} & Cosmological Zoom & $z=0$ Disk & {\sc gizmo} HDF5 & \footnote{\url{https://fire.northwestern.edu/}} &\citet{hopkins18a} \\
                & & & & & \citet{wetzel16a}\\
                {\sc GasolineDisk} & Idealized & Isolated Disk & {\sc gasoline} TIPSY& \footnote{\url{http://yt-project.org/data/TipsyGalaxy.tar.gz}} & \\
                {\sc ChangaMW} & Cosmological Zoom & $z=0$ Disk & {\sc changa TIPSY} &  \footnote{\url{https://users.astro.ufl.edu/~desika.narayanan/powderday_files/changa_starform_example}} & \citet{sanchez19a}\\
                &&&&&\citet{tremmel17a}\\
                {\sc SmuggleDisk}&  Idealized & Isolated Disk & {\sc arepo} HDF5 & \footnote{\url{https://users.astro.ufl.edu/~desika.narayanan/powderday_files/smuggle_snapshot_143.low_res.hdf5}}&\citet{marinacci19a} \\
                {\sc tnghalo} & Cosmological & Galaxy Cluster& {\sc arepo} HDF5&\footnote{\url{http://yt-project.org/data/TNGHalo.tar.gz}} &\citet{pillepich18a} \\
                {\sc EnzoDisk} & Idealized & Isolated Disk & {\sc enzo} & \footnote{\url{http://yt-project.org/data/IsolatedGalaxy.tar.gz}} & \citet{kim14a}\\
                {\sc simba m25n512} & Cosmological &N/A & {\sc gizmo} HDF5 & \footnote{Available by request} & \citet{dave19a}\\
        \end{tabular}
\end{table*}

\subsection{Test Model Galaxies}
\label{section:galaxymodels}
Throughout this paper, we will provide both model tests and examples
of the code's capabilities on a number of different simulation
datasets.  In Table~\ref{table:models}, we summarize these models.  In
summary, we seek to use a diverse range of hydrodynamic simulation
codes as well as simulation types (i.e. idealized, cosmological
zoom-in, and bona fide cosmological).  We use these throughout this paper in various tests and examples in part to demonstrate the seamlessness with which \pd \ interfaces with a diverse range of galaxy formation models.  We describe these models briefly here.

\begin{enumerate}
\item {\sc gizmodisk} is a cosmological zoom-in simulation of a
  disk-like galaxy at $z \approx 0$ run by
  \citet{narayanan18a,narayanan18b,li18a,privon18a}. This simulation was run with the
  hydrodynamic code {\sc gizmo}, with the {\sc mufasa} suite of
  galaxy formation physics enabled \citep{dave16a}.

    \item {\sc latte} is a cosmological-zoom in simulation of a
        Milky Way-like galaxy from the Latte simulation series.  The
        Latte suite of FIRE-2 cosmological zoom-in baryonic
        simulations of Milky Way-mass galaxies \citep{wetzel16a}, part
        of the Feedback In Realistic Environments (FIRE) simulation
        project, were run using the {\sc gizmo} gravity plus hydrodynamics
        code in meshless finite-mass (MFM) mode \citep{hopkins15a} and
        the FIRE-2 physics model \citep{hopkins18a}.

  \item {\sc GasolineDisk} is an idealized {\sc gasoline} simulation of a
    disk-like galaxy, publicly available at \url{https://yt-project.org/data/}.

      \item {\sc changamw} is a cosmological zoom-in simulation of a Milky Way mass galaxy at $z \sim 0$ performed by \citet{tremmel17a} and \citet{sanchez19a} with the {\sc changa} hydrodynamic code.

    \item {\sc SmuggleDisk} is an idealized Milky Way-like disk galaxy
      run with the {\sc arepo} hydrodynamic code by \citet{marinacci19a}, with the {\sc smuggle} physics suite enabled. 
      
        \item {\sc tnghalo} is an {\sc arepo} simulation of a massive halo from the {\sc illustris-TNG} simulation, and is publicly available in snapshot form at \url{https://yt-project.org/data/}.

      \item {\sc EnzoDisk} is an idealized disk galaxy run with {\sc enzo}, and publicly available at \url{https://yt-project.org/data/}.

        \item {\sc simba m25n512} is a cosmological simulation first
          run for the study of \citet{narayanan18b}. This cosmological
          box employing the {\sc simba} galaxy formation physics model
          \citep{dave19a,li19a} is $25/h$ Mpc on a side.
          
\end{enumerate}

\subsection{Grid Construction and Structure}

Capitalizing on the flexibility afforded by {\sc hyperion}, \pd \ is
able to perform radiative transfer for hydrodynamic simulations that
are particle-based, operate on adaptive meshes, and on unstructured
meshes.

  The radiative transfer happens on a mesh.  For particle-based codes, the physical  properties of the particles are
  projected onto an adaptive mesh with a octree memory structure, and
  smoothed utilizing a Spline kernel.  The hierarchy in the octree is
  depth-first.  Formally, the entire data set of particles is
  encapsulated into a single cell, which then recursively refines into
  octs until a maximum threshold number of particles are present
  in a cell.  
 

For codes that operate on a Voronoi mesh (i.e. {\sc arepo}), \pd
\ leverages \hyperion's ability to perform radiative transfer on the
mesh itself, and re-constructs the mesh based on the particle
positions.  Similarly, for adaptive mesh codes  (e.g. {\sc enzo}), \pd
\ performs the radiative transfer directly on the grid used for the
evolution of fluid quantities in the hydrodynamic simulation.

     \subsection{Stellar Population Synthesis}

      \begin{figure}
\includegraphics[scale=0.55]{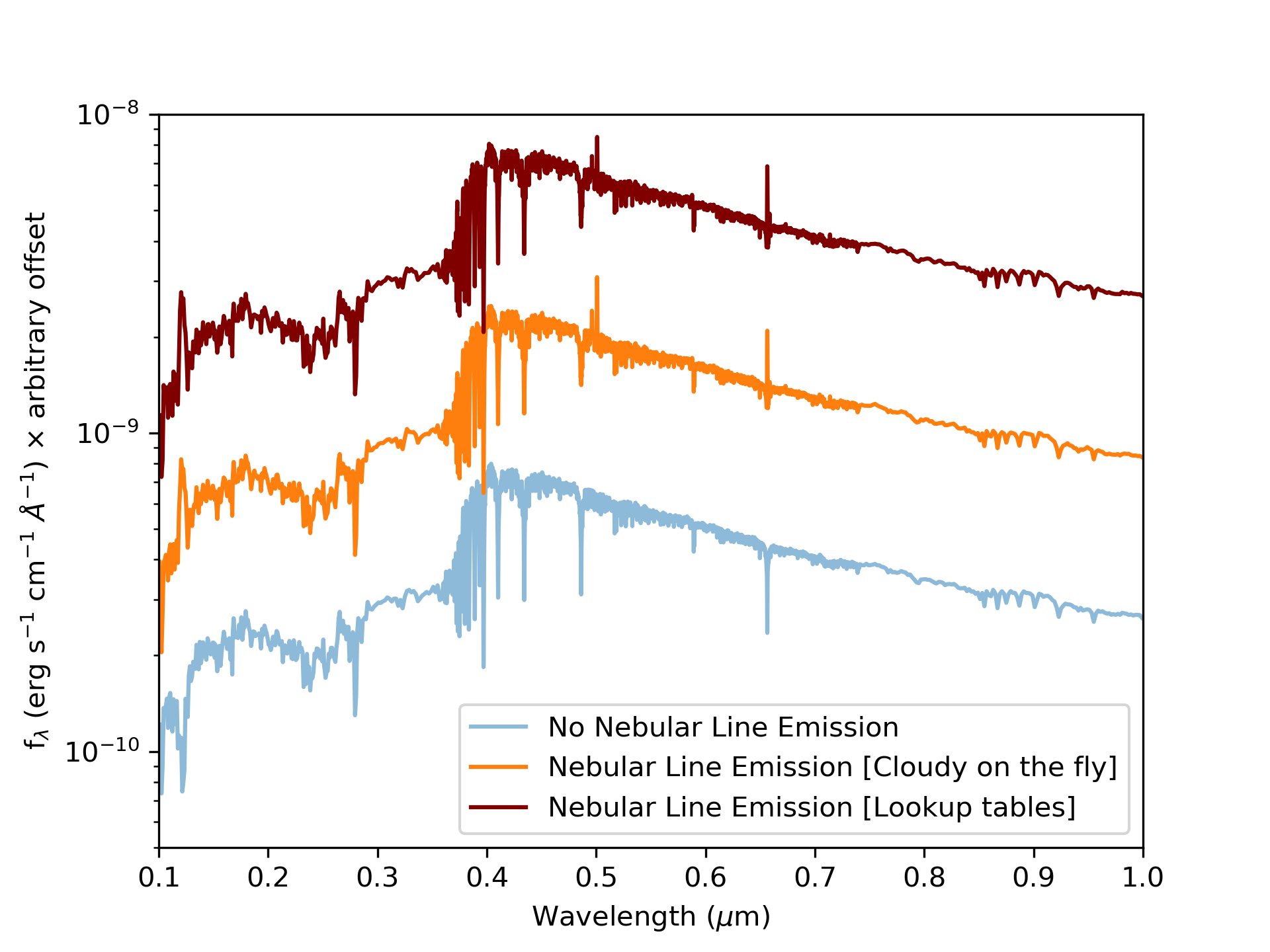}
\caption{Impact of nebular line emission on UV-optical SED of a
  star-forming galaxy (model {\sc GizmoDisk}).  The blue line shows
  the default model with no nebular line emission; the orange line shows the UV-optical SED
  (including nebular lines) for a model in which the spectrum from photoionization regions
  around young stars are calculated with on the fly {\sc cloudy} models, while the maroon line
  shows the same, but with the nebular line emission computed via the \citet{byler17a} lookup
  tables.   The flux densities (ordinate) are offset by an arbitrary multiplicative to aid in clarity.
  \label{figure:nebular} }
\end{figure}

The stellar clusters in simulations emit SEDs based on their metallicities and
ages which are drawn from the galaxy simulations\footnote{Sometimes a
  population of ``old stars'' are initialized with idealized
  simulations.  For these stars, the ages and metallicities are input as a
  free parameter.}.  To calculate these, we leverage the high level of
flexibility available in the \fsps \ population synthesis
code\footnote{On a practical level, in order to interface with the
  Fortran-based \fsps, we utilize the publicly available \fsps
  \ python hooks originally developed by D. Forman-Mackey
  (\url{http://dan.iel.fm/python-fsps}) }
\citep{conroy09b,conroy10a,conroy10b}.  The methodology of
constructing a stellar population synthesis is covered in the
aforementioned papers, as well as the reviews by \citet{walcher11a,conroy13a}, and
we refer the reader to these works for detailed discussion.   

This level of flexibility adds a powerful dimension to \pd \ currently
unavailable in any other public dust radiative transfer package.  In
principle, nearly any population synthesis option available in \fsps
\ is also available for variation in \pd.  In practice, the default
version of the code ships with the ability to handle variable
functional forms for the stellar IMF (with both relatively standard
options available such as \citet{salpeter55a}, \citet{chabrier03a},
and \citet{kroupa02a}, as well as user-specified IMFs), a range of theoretical isochrones  \citep[e.g.][]{bertelli94a,pietriferni04a,schaller92a}, varying
contributions to the SED from post AGB stars, a circumstellar AGB dust
model, obscuration of young stars by unresolved birth clouds, and
nebular line emission \citep[as we will discuss in \S~\ref{section:nebular}, both building off of the \fsps \ libraries
  developed by ][as well as via direct {\sc cloudy} modeling]{byler17a,byler18a,byler19a}.  A key point here is
that because \fsps \ is actively being developed and maintained, new
features developed in this population synthesis code will also be available in \pd.

\subsection{Nebular Line Emission}
\label{section:nebular}
\pd \ includes nebular line emission from \ion{H}{2} regions around massive
stars using {\sc cloudy} calculations.  These come in two flavors:
lookup tables (that are relatively efficient), and slower but more
flexible direct {\sc cloudy} models that are run on the fly.  Because
both methods tie the nebular line emission to the star particles
themselves, these lines are attenuated by any diffuse dust they see as
they exit the galaxy.

The first method uses {\sc cloudy} lookup tables generated for {\fsps}
stellar population synthesis models developed by 
\citet{byler17a,byler18a,byler19a}.  These lookup tables, computed
with {\sc cloudy v13.03} are built for a grid of stellar age ($t_{\rm age}$), metallicity ($Z$) and ionization parameter  which
range from $-4 \leq {\rm log_{\rm 10} \ U} \leq -1$, $-1.98 \leq {\rm
  log_{\rm 10} \ Z_{\rm HII}} \leq 0.198$, and $0.5 \leq t_{\rm age}
\leq 20$ Myr.  Here, the ionization parameter is the usual
dimensionless ratio between the number of ionizing photons and
hydrogen density:
\begin{equation}
  \label{equation:logu}
U_0 \equiv \frac{Q_{\rm H}}{4\pi R_{\rm HII}^2 \times n_{\rm H,HII} \times c}
\end{equation}
where $Q_{\rm H}$ are the total number of hydrogen ionizing photons emitted per second:
\begin{equation}
Q_{\rm H} \equiv \frac{1}{hc}\int_0^{912 {\rm \AA}} \lambda f_\lambda d\lambda
\end{equation}
$n_{\rm H,HII}$ is the density of the \ion{H}{2} region, and is assumed to be
fixed at $n_{\rm H,HII} = 100$ cm$^{-3}$.  $R_{\rm HII}$ is the radius
of the \ion{H}{2} region.  The formal definition for $U$ uses the Str\o mgren
radius.  However, this is only known after the photoionization state
is computed!  Therefore, $R_{\rm HII}$ is set to $R_{\rm HII}=R_{\rm
  inner,HII}$, which is the inner boundary of the \ion{H}{2} region, and the
quantity of interest for the {\sc cloudy}
calculations. \citet{byler17a} assume $R_{\rm inner} = 10^{19}$ cm.

\pd \ additionally includes a number of options relevant to nebular
line emission from \ion{H}{2} region relevant to galaxy-wide radiative
transfer. First, many simulations (especially cosmological ones) have
mass resolutions that are significantly larger than the mass of a
typical stellar cluster.  This can lead to unphysically large
ionization parameters, $U_0$ owing to the increased number of
Lyman-limit photons.  We therefore allow the user to subdivide stellar particles into a mass spectrum of stellar clusters following a powerlaw function:
\begin{equation}
\frac{dN}{dM} \propto M^{\beta}
\end{equation}
based on observational constraints by \citet{chandar14a,chandar16a}.
Each of these clusters then radiates its own individual SED, though
are assumed to be cospatial at the point of the parent star particle.  Second, while $U,Q$ and the metallicity of the \ion{H}{2} region ($Z_{\rm
  HII}$) are all calculated based on the particle properties, it is
conceivable that the user may wish to hold these fixed as an assumed
value, and can therefore be set by the user. 
Alongside lookup tables, {\sc powderday} allows for the direct computation on the fly of nebular line emission from all stars (or stars within certain age thresholds for computational ease).  For these, we couple the simulations to {\sc cloudy} \citep{ferland13a}.
  This offers
significant advantages over the aforementioned lookup table-dependent
methods as it obviates the user having to generate new lookup tables
for every new set of assumed stellar parameters.  For these
calculations, we assume a spherical \ion{H}{2} region geometry in which the
inner boundary of the \ion{H}{2} region is set to be the Str\o mgren radius:
\begin{equation}
R_{\rm S} = \left(\frac{3Q_{\rm H}}{4\pi n_{\rm H}^2 \alpha_{\rm B}}\right)^{1/3}
\end{equation}
where the $n_{\rm H}$ has a default value of $100 $ cm$^{-3}$, and the
temperature of the region has a default value of $T=10^4$ K for the
calculation of $Q_{\rm H}$ (though both are adjustable).

While the direct calculation of emission from \ion{H}{2} regions on a
particle-by-particle basis can be slow, it offers two distinct
advantages over the lookup tables.  First, there is a significant
flexibility advantage.  For example, if a user wants to include dust
in \ion{H}{2} regions using lookup tables, they would have to completely
regenerate the lookup table.  When employing {\sc cloudy} on the fly,
it is straightforward to simply update this in the {\sc cloudy}
parameter file and re-run the {\sc powderday} simulation.  Second, as
previously mentioned, the lookup tables are parameterized in a grid of
stellar age, metallicity, and ionization parameter, where the
resulting nebular line emission for a given star particle is
interpolated within this grid.  By employing the direct {\sc cloudy}
simulations on a particle-by-particle basis, one is able to avoid this
interpolation, which can impact the expected fluxes from individual
lines.

In Figure~\ref{figure:nebular}, we show an example of the nebular line
models in our simulations.  We show the UV-optical SED of model galaxy
{\sc GizmoDisk} in three cases: with nebular line emission turned off,
with the nebular line emission calculated via the \citet{byler17a}
lookup tables, and via direct {\sc cloudy} calculations.  The largest
impact to the UV-optical SED obviously occurs when comparing a model
with no emission from \ion{H}{2} regions vs a model that includes it: the
addition of \ion{H}{2} regions to the source term contributes to both
continuum and line emission in the UV/optical.  While the model using
lookup tables is fairly similar to that using {\sc cloudy} on the fly,
there are of course quantifiable differences in the line and continuum
strength.  These primarily owe to the interpolation in (ionization
parameter; stellar age; metallicity) space for the lookup tables,
versus direct calculation.

\subsection{Active Galactic Nuclei}
Radiation from accreting black holes can also be included in \pd.  We
assume that the luminosity of the black hole is proportional to the
mass accretion rate, modulated by an efficiency parameter $\eta$:
\begin{equation}
  \label{equation:l_agn}
  L_{\rm AGN} = \eta \dot{M}_{\rm BH} c^2
\end{equation}
Where $L_{\rm AGN}$ is the black hole luminosity, $\dot{M_{\rm BH}}$ is the black hole accretion rate, and $c$ is the speed of light.  The template spectrum for the AGN are based on the
luminosity-dependent templates of observed unreddened type 1 quasars
\citep{hopkins07a}.  Hot dust emission from the putative torus is
included in the template, and is based on the mid-IR SED template of
\citet{richards06a}.  The normalization of this template is based on
the bolometric luminosity for the AGN, given by
Equation~\ref{equation:l_agn}, and the efficiency $\eta$ is a free
parameter.

As an alternative to the \citet{hopkins07a} AGN SED templates, \pd
\ also includes the option to employ the {\sc clumpy} SED templates
from \citet{nenkova08a,nenkova08b}. {\sc clumpy} models an inhomogeneous dust obscuring structure around the AGN and provides a grid of SED templates that include torus-scale dust obscuration and emission. The dust structure is fully described by the dust optical depth $(\tau_V)$, torus inclination $(i)$, number of dust clumps along the line of sight $(N_0)$, and the angular and radial spatial distributions of the clumps $(Y, q, \sigma)$. The dust grains are assumed to have optical properties of a standard Milky Way mixture, which \citet{nenkova08b} find agree with observed AGN dust compositions. Clumpy dust structures provide a potential explanation for the observed behavior of 10$\mu$m silicate absorption in AGN \citep{mason06}, as well as the compact size of the torus \citep{poncelet06}, the close spatial proximity of vastly different dust temperatures within AGN \citep{schartmann05}, and broad-line emission at equatorial viewing angles \citep{alonso-herrero03a}.



\begin{figure}
  \includegraphics[scale=0.55]{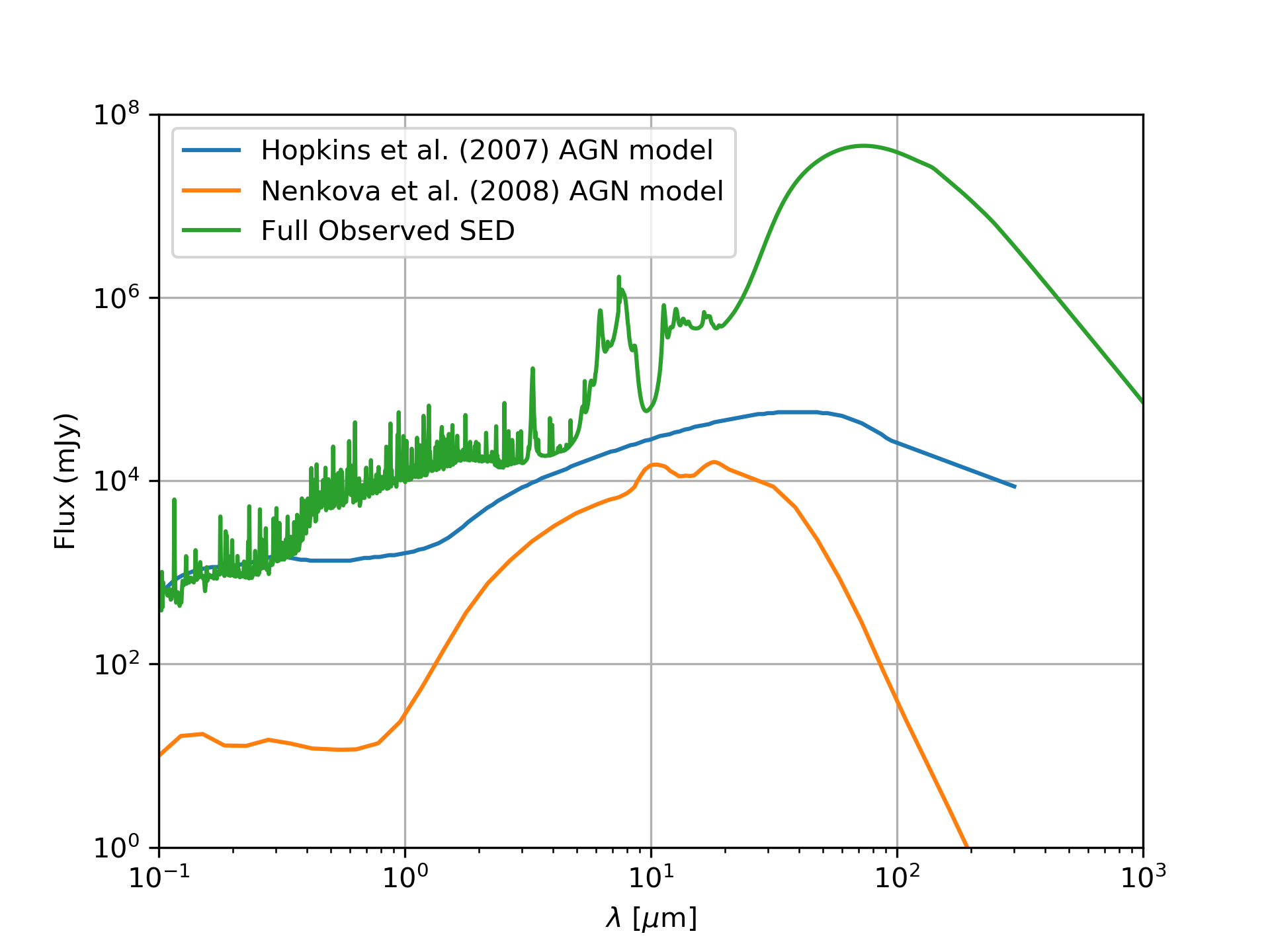}
  \caption{Example of possible input SEDs for accreting black holes.
    The blue line shows the \citet{hopkins07a} template SED, while the
    orange line shows the \citet{nenkova08b} model.  The green line
    shows the output observed SED (including the contribution from
    stars and dust).\label{figure:bh_sed}}
\end{figure}

In Figure~\ref{figure:bh_sed}, we show an example model demonstrating how
the \citet{hopkins07a} and \citet{nenkova08b} AGN models are included.
The modeled galaxy is an arbitrarily chosen one (galaxy \#12) from the
{\sc simba} m25n512 simulation.  The solid green line is the output
observed SED, while the blue and orange lines show the input model AGN
SEDs (of course in a realistic run only a single model would be
employed at a time).   For the
\citet{nenkova08b} template, we assume a default set of parameters of
      [$N_0,Y,i,q,\sigma,\tau_{\rm V}$] = [$5,30,0,1.5,30,40$].

Finally, \pd \ includes the option to apply post-processed, short-timescale AGN variability using the analytic prescription from \citet{hickox14a}. This prescription gives the relative time, $t$, spent by an AGN at a given fraction of its bolometric luminosity, $L_{\rm rel}$. It takes the form of a Schechter function with an exponential cutoff at $L_{\rm cut} = 100 L_{\rm AGN}$ and a lower limit of $10^{-5} L_{\rm AGN}$:

\begin{equation}
    \label{equation:l_agn_variability}
    \frac{dt}{d\log L_{\rm rel}} = t_0 \left(\frac{L_{\rm rel}}{L_{\rm cut}}\right)^{-\alpha} \exp\left(-L_{\rm rel}/L_{\rm cut}\right),
\end{equation}
where the characteristic timescale, $t_0$, is adjusted such that the integral over all $L_{\rm AGN}$ is $1$. We use the \citet{hickox14a} fiducial model with power-law slope $\alpha = 0.2$, which they find gives a robust compromise between observed Eddington ratio distributions \citep{hopkins09f,kauffmann09} and simulated AGN variability \citep{novak11}. \citeauthor{hickox14a} find that applying short-timescale variability to observations reproduces general trends in AGN luminosity functions and merger fractions, and yields a close connection between AGN activity and star formation rates over galaxy evolution timescales. For each simulation snapshot, we sample the prescribed luminosity distribution, vary the black hole bolometric luminosity according to the drawn relative luminosity, then continue the radiative transfer.

\subsection{Cosmic Microwave Background}
The cosmic microwave background (CMB) is included as an additional
energy density term in every cell in the simulation.  Specifically, we model this as:
\begin{equation}
  \epsilon = \int \kappa_\nu B_\nu d\nu \ \rm{erg\ s^{-1} g^{-1}}
\end{equation}
where $\kappa_\nu$ is the dust absorption opacity (based on the
assumed extinction properties of the dust grains) and $B_\nu$ is the
Planck function.  As demonstrated by \citet*{privon18a}, this heating
term can be non-negligible for high-redshift ($z\ga4-5$) galaxies.

\subsection{Dust Content}
\label{section:dust_models}
For all types of grids, the fundamental quantity of interest for the
radiative transfer is the dust density, which can be specified in a
number of manners.  To specify the dust content in a given grid cell,
we include both observationally-motivated and theoretically-motivated
methods for determining the dust mass.  The simplest and most traditional method for
determining the dust mass is to employ a constant dust mass to metals
mass ratio.  Indeed, a relatively narrow range of values has been
reported by a number of authors over a diverse range of galaxy
environments and redshifts
\citep[e.g.][]{dwek98a,vladilo98a,watson11a}. Alternatively, recent
observations by \citet{remyruyer14a} and \citet{devis19a} have
demonstrated a trend between the dust to gas ratio and metallicity of
galaxies.  Accordingly, we include this scaling \citep[specifically,
  the best fit single power-law relation by][in which the CO-H$_2$ conversion factor is allowed to vary with metallicity]{remyruyer14a}. 

  Similarly, advances in galaxy formation algorithms in the last few
  years have ushered in a new suite of models that include on-the-fly
  dust formation, growth, and destruction processes
  \citep[e.g.][]{asano13a,mckinnon16a,mckinnon18a,popping17b,aoyama17a,aoyama18a,hou17a,hou19a,li19a}.
  For these types of simulations, \pd \ can explicitly read in the
  dust masses from the simulation themselves, offering
  self-consistency with the galaxy formation simulation.

  Finally, we include the option of generating dust masses by
    leveraging the capabilities of simulations that include on the fly
    dust physics, even for galaxy models that do not include dust
    physics.  To do this, we employ the results of \citet*{li19a},
    which uses the {\sc simba} dust formation, growth and destruction
    framework to map the physical properties of galaxies to their dust
    content.  We provide two options from the \citet*{li19a} model.
    The first is an approximate mapping between the dust to gas ratio
    (DGR) and the gas phase metallicity:
    \begin{equation}
      \label{equation:li_dgr}
      {\rm log \ DGR} = 2.445 \times \left(\frac{Z}{Z_\odot}\right)-2.029
    \end{equation}
    This relation carries two sources of uncertainty with it.  First,
    similar to the observational work of \citet{remyruyer14a}, which
    reports a similar mapping, Equation~\ref{equation:li_dgr} is
    constructed for galaxy-wide scales, which therefore provides a
    similar uncertainty as employing the \citet{remyruyer14a} relations
    when applying these relations to resolved scales within galaxies (i.e. on a particle-by-particle or cell-by-cell basis).
    The second uncertainty folded into Equation~\ref{equation:li_dgr}
    is that it there is significant scatter associated in this
    mapping. A primary result of \citet*{li19a} was that there are numerous secondary
    dependencies between the dust to gas ratio and physical properties
    of galaxies beyond the gas phase metallicity that, when included
    in the mapping, can significantly reduce the scatter.

    To move beyond these two sources of uncertainty,
      \citet*{li19a} developed a machine learning framework to map
      between the DGR of galaxies and their physical properties,
      thereby reducing the scatter intrinsic in the single parameter
      mapping between DGR and $Z$.  To inform dust mass calculations
      on resolved scales for \pd, we build on the \citet*{li19a} model,
      and provide a mapping between the DGR and metallicity, star
      formation rate, and gas mass of every particle in the {\sc
        simba-100} (100 Mpc)$^3$ cosmological simulation at redshift
      $z=0$.  This mapping utilizes the Extremely Randomized Trees
      method within the {\sc scikit-learn} software package.  The
      advantage of this model is that it allows modelers who do not
      otherwise have information about the dust content of their
      galaxy to take advantage of results from simulations that do,
      thereby allowing for increased sophistication over more typical
      constant dust-to-metals ratio assumptions.

\begin{figure}
        \includegraphics[scale=0.55]{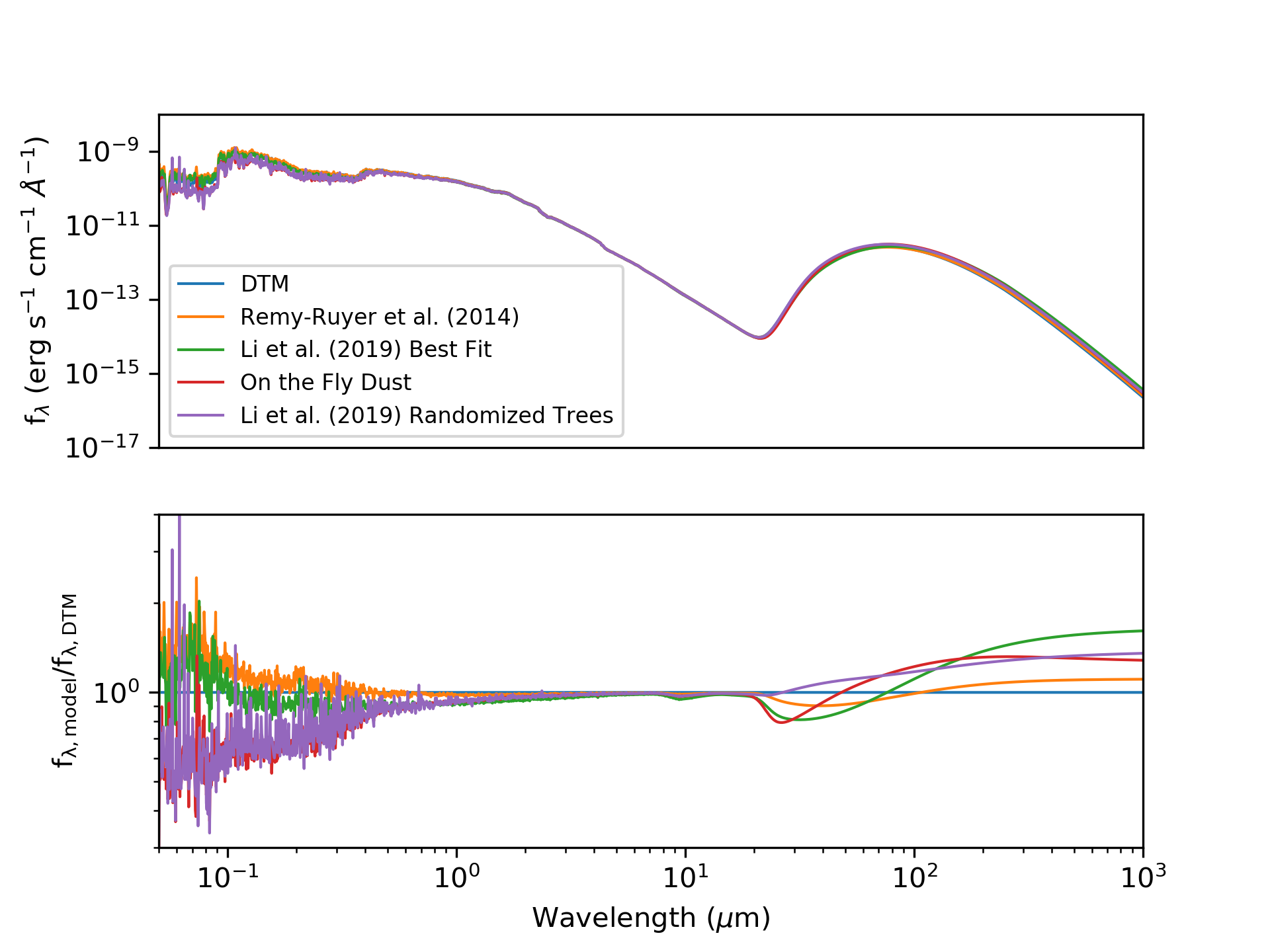}
        \caption{Impact of different choices for modeling dust on
          derived SEDs.  We investigate modeling the dust via a simple
          dust to metals ratio (DTM), following the
          \citet{remyruyer14a} observational scaling with metallicity
          on galaxy-wide scales, using the \citet{li19a} best fit
          relation between the dust to gas ratio and metallicity from
          simulated galaxies, an explicit on-the-fly dust calculation,
          and via the \citet{li19a} machine learning framework.  In
          the top panel we show the actual SEDs from these models, and
          in the bottom panel the relative errors.  The model employed
          here was {\sc Galaxy9} from the $z=0$ snapshot of the m25n512
          {\sc simba} simulation.\label{figure:dust_model_comparison}
         }
      \end{figure}

      In Figure~\ref{figure:dust_model_comparison}, we show the impact
      of these different dust model choices on a calculated SED (from
      galaxy 9 in the $z=0$ snapshot of the {\sc simba} m25n512
      simulation).  It is important to note that this model is simply
      an example (and indeed the galaxy was chosen arbitrarily), and
      not representative of all model galaxies.  With this in mind, we
      see that the biggest impact is on both the UV radiation, as well
      as the mid-FIR SED.  Factors of $\sim 2-4$ differences in the
      attenuated UV flux alone can be ascribed solely to the
      implemented dust model.

      \begin{figure}
\includegraphics[scale=0.55]{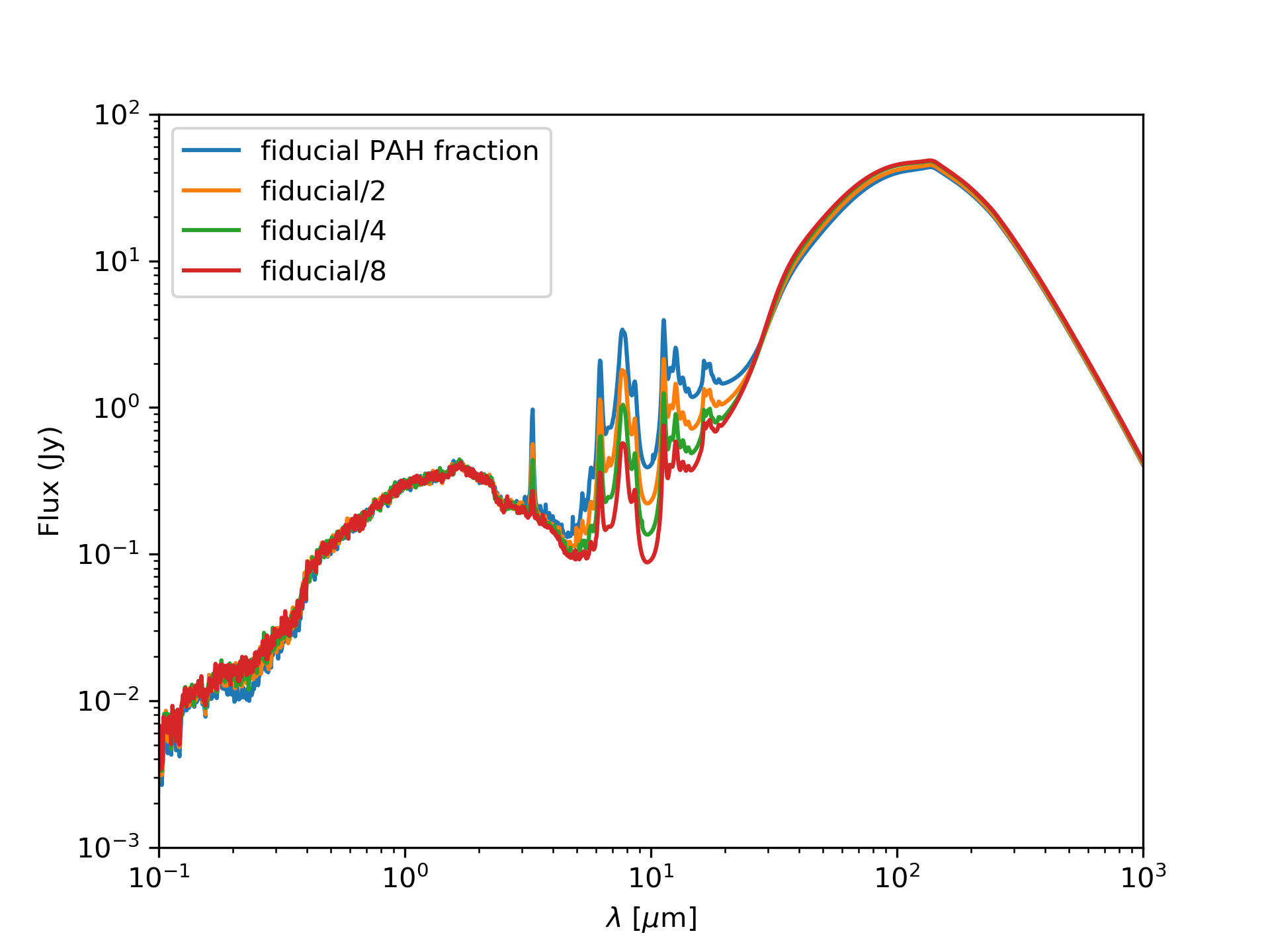}
\caption{Model SED from example galaxy {\sc latte} with decreasing
  ultra small grain (USG) mass fraction.  The reduced fractions of
  USGs modify both the UV continuum, as well as the PAH intensity.
  See text for details. \label{figure:pah}}
\end{figure}

      \subsection{Polycyclic Aromatic Hydrocarbons}
      \label{section:pahs}
 We
      follow the methodology of \citet{robitaille12a} for including
      polycyclic aromatic hydrocarbons (PAHs) as they are modeled in
      {\sc hyperion}.  The PAH model is based on a modified version of
      the \citet{draine07a} model.  We utilize the \citet{draine07a}
      emissivities and opacities for dust grains, though bin the
      grains into three size distributions: ultra small grains (USGs;
      $a < 20 {\rm \AA}$), very small grains (VSGs; $20 {\rm \AA} < a < 200 {\rm \AA}$)
      and big grains ($a > 200 {\rm \AA}$).  Here, the PAHs are assumed to
      be exclusively in the smallest (USG) bin, while the grains in
      the largest bin follow the adopted global grain size
      distribution (i.e. \citet{weingartner01a}.  The distribution of
      USGs, VSGs and big grains can be set by the user, though have a
      default proportion of ($5.86\%, 13.51\%$  and $80.63\%$)
      respectively.

      Traditionally, the dust emissivities in the \citet{draine07a}
      formalism are computed for variable radiation intensities,
      scaled by the interstellar radiation field in the solar
      neighborhood as computed by \citet{mathis83a}.  As discussed in
      \citet{robitaille12a}, {\sc hyperion} instead parameterizes the
      radiation field by the power absorbed by grains, which accounts
      for differing spectral shapes.  The dust emissivities are
      computed for bins of power of radiation field absorbed by the
      grains per unit mass:
      \begin{equation}
        \dot{A} = \int 4\pi J_{\rm isrf} \kappa_\nu d\nu
      \end{equation}
      where $J_{\rm isrf}$ is the mean intensity of the radiation
      field at the location of the dust grain.

      In Figure~\ref{figure:pah}, we demonstrate the impact of the USG
      fraction on the PAH emission intensity via an example SED (model
      {\sc latte}) in which we run with our default USG/VSG/big grain
      setup, as well as by incrementally halving the USG mass
      fraction. As the USG fraction decreases, the UV flux density
      increases as the power in the PAHs commensurately decreases.
      Currently, this mass fraction of grains in PAHs is a
      free parameter, though future versions of {\sc powderday} will
      have the capability to include an actively modeled size spectrum
      of dust grains \citep[e.g.][Qi Li et al., in prep.]{asano13a,nozawa15a,hirashita15a,hirashita19a,gjergo18a,aoyama18a,mckinnon18a,hou19a}

\subsection{Dust Radiative Transfer}

The radiation emitted from luminous sources are then allowed to
propagate through the dusty ISM, where the dust masses are computed as
described in \S~\ref{section:dust_models}.  These photons propagate
through the grid cells, and can be scattered, absorbed (and re-emitted), or pass
through freely.  This continues until the photons leave the grid.

We utilize \hyperion \ as the central dust radiative transfer solver
\citep{robitaille11a}.  \hyperion \ is an ultra-flexible code that
solves for the transfer in a Monte Carlo fashion, and utilizes the
\citet{lucy99a} iterative method for determining the equilibrium dust
temperature.  We note that in what follows, we describe \hyperion \ as
it is used as a part of \pd.  The code contains significantly more
options and nuance than is described here, and we refer the reader to
\citet{robitaille11a} for a more detailed description. 

We add sources as point sources, with a given SED shape and
luminosity.  In order to reduce memory overhead, we bin the sources in
age and metallicity.  Without such a procedure, adding all of the
stellar clusters from even a relatively low mass-resolution galaxy
simulation would be prohibitive.  When included, black holes are added
in the same manner as stellar clusters.  Photons are randomly sampled
from sources, with numbers proportional to the source luminosity.  The
direction and frequency are randomly drawn, and the photon is
propagated until it either escapes the grid or reaches an arbitrary
optical depth, $\tau$, where $\tau$ is randomly drawn from the
exponential distribution $e^{-\tau}$.  Formally, $\tau = -{\rm
  ln}\xi$, where $\xi = [0,1]$.  Whether the photon is absorbed or
scattered at this point is dependent on the dust albedo.  

The equilibrium dust temperature\footnote{In practice what is
  calculated is the equilibrium dust emissivity and mean opacity,
  which are functions of temperature.} is calculated iteratively until
convergence.  This is because the emissivity depends on the mean
radiation field which depends on the emissivity.  The iteration
continues until a threshold number of cells have differences in
specific energy absorption rates below a defined value, and their
values have changed by less than a relative threshold value.


SEDs and images are calculated via ray tracing.  The source function
is determined at each position in the grid, and then the 
radiative transfer equation is integrated along lines of sight.  The SEDs are
comprised of wavelengths where radiation is emitted (though bounded by
the wavelengths of dust opacity tables).  Images are made only at
pre-specified wavelengths in order to save on memory cost.

\begin{figure}
  \includegraphics[scale=0.55]{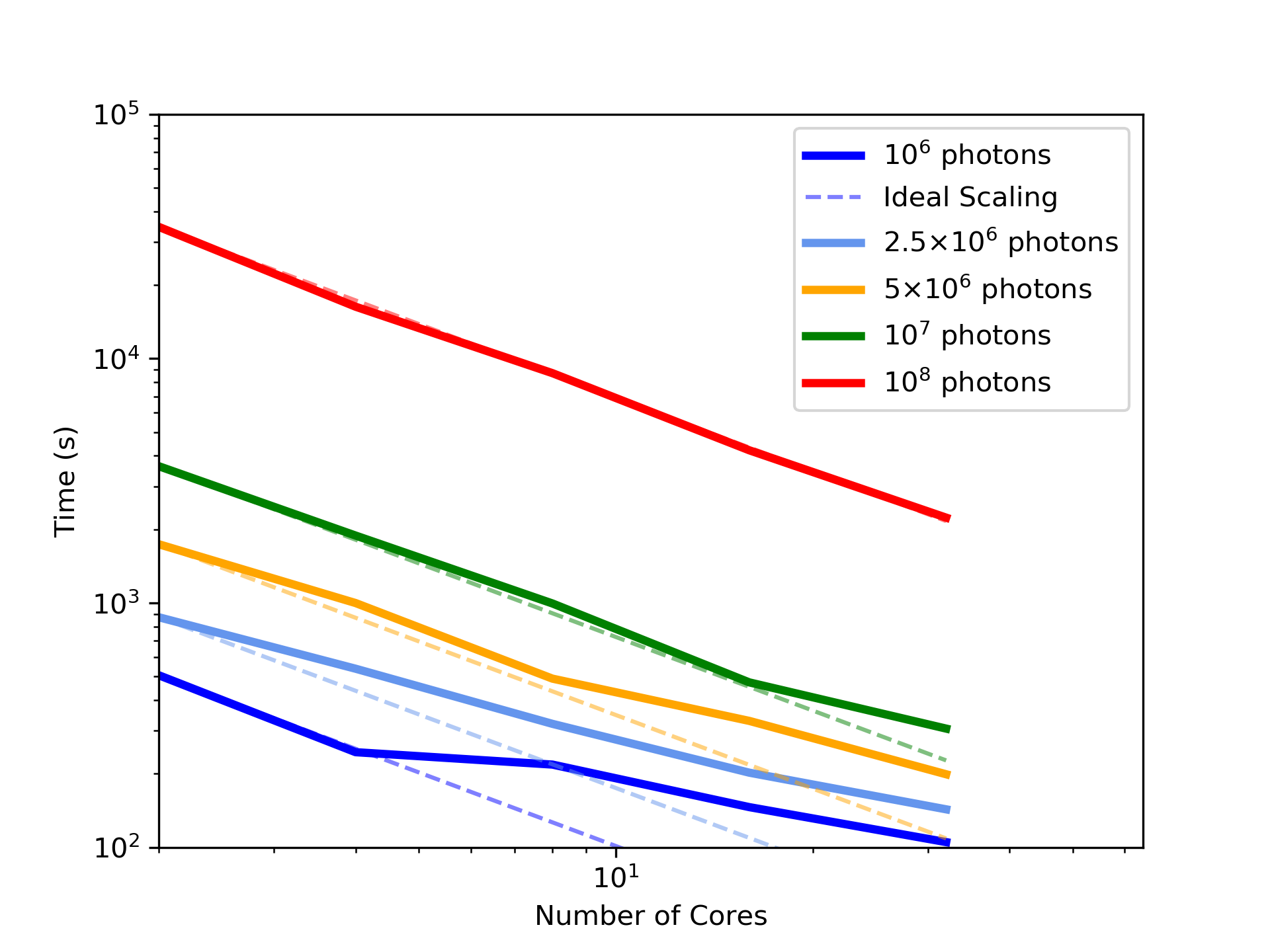}
  \caption{Scaling tests for \pd \ presented as wall clock time as a function of
    number of cores for problems of increasing difficulty
    (parameterized by the total number of photons emitted).  \pd \ generally shows strong scalability, though at low photon-count (i.e., easy) problems, the fixed overhead costs in the pre-radiative transfer stage can drive some inefficiencies.  \label{figure:scaling}}
\end{figure}

\begin{figure*}
  \includegraphics{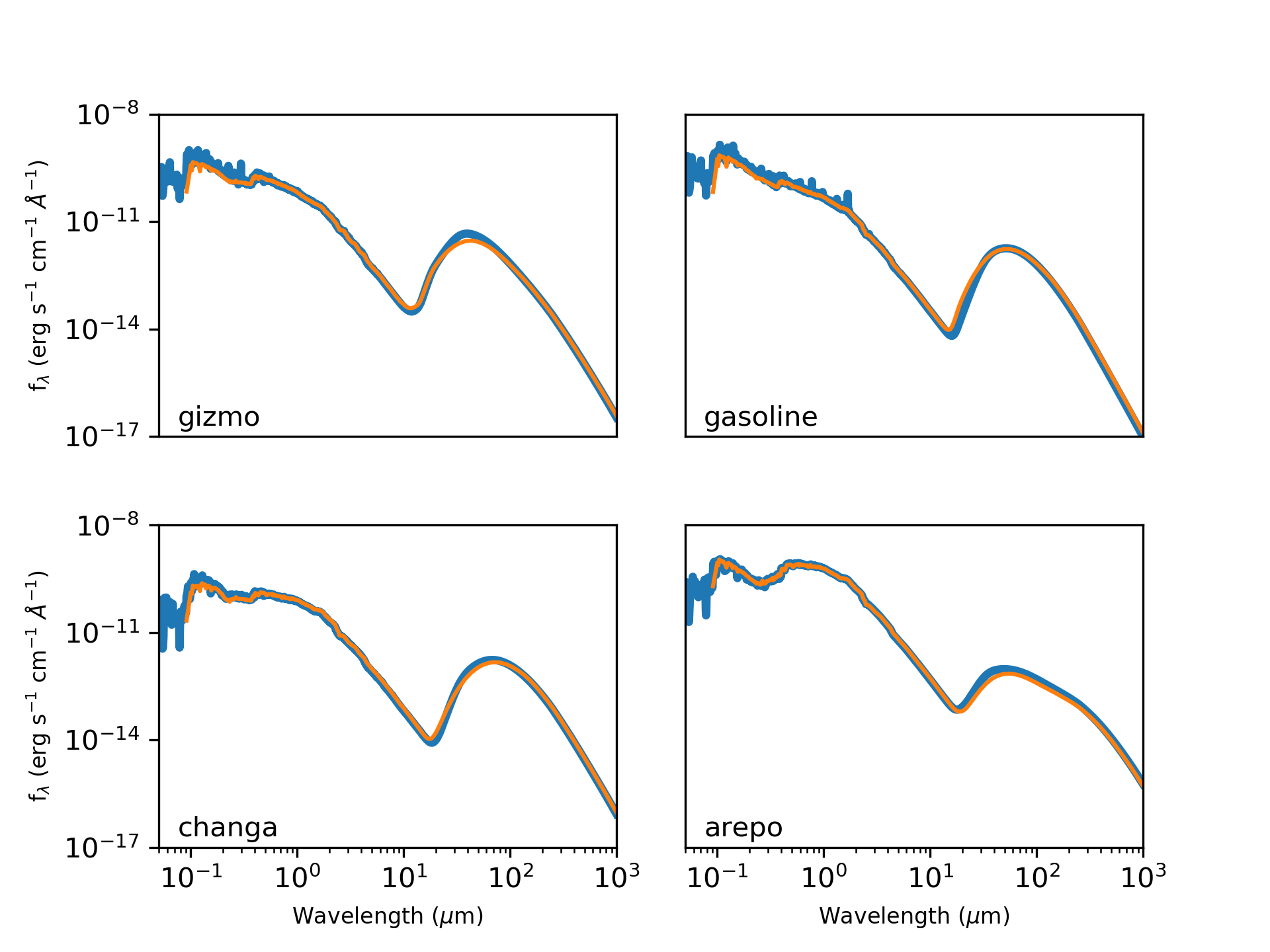}
  \caption{Model comparison between \pd \ and \skirt \ for {\sc
      gizmo}, {\sc gasoline}, {\sc changa}, and {\sc arepo} simulations.  
    While there are differences in the inherent stellar SED models, as
    well as the grid construction, the high level of agreement between the output of the two
    simulation codes is
    striking. \label{figure:powderday_skirt_comparison}}
\end{figure*}

\section{Code Description -- Implementation}
\label{section:implementation}

\subsection{Front Ends}
In order to aid in user-ease, \pd \ leans on \yt \ for reading in
galaxy snapshots.  \pd \ reads in individual 
 snapshots as a \yt \ dataset object, and therefore has
all of the associated methods and attributes offered in \yt \ available to it.  The
simulation type is automatically detected within \yt, and passed to
the appropriate \pd \ front end which converts the unique field names and
units associated with each simulation type to internal ones so that
downstream from the front end all physics in the code remains the
same, regardless of the input hydrodynamic simulation.  Currently,
front ends exist for {\sc gizmo}, {\sc gasoline}, {\sc
  changa}, {\sc arepo} and {\sc enzo}.

\subsection{Code Scaling}
\pd \ is a parallelized code that offers reasonable scaling with
processor number.  The code is fundamentally broken into two regimes:
the stellar population synthesis and model setup, and the
radiative transfer.  The most costly aspect of the model setup is the SED generation.  As a result, this is parallelized via the {\sc
  pool.map} multithreading package in Python.  The radiative
transfer in {\sc hyperion} is fully MPI parallelized.  What this means
is that there is a fixed overhead for a given problem in generating
the stellar population and other model setup procedures, while the radiative transfer
itself represents a highly scalable parallel problem.  How the problem scales will depend in part on the computer configuration.  The SED generation can operate on many tasks on a single compute node, whereas the radiative transfer can employ multiple networked nodes.  As a result, the optimal configuration for parallelization is to employ many tasks on a single compute node:  this allows for both the stellar SED generation and the radiative transfer to be maximally parallelized.  Spreading a given {\sc powderday} simulation across many nodes will lose some efficiency as the initial SED generation will still only occur on a single node.  

In Figure~\ref{figure:scaling}, we present the
result of a scaling test with model {\sc SmuggleDisk}, in which we
increase both the number of processors, as well as the number of
photons.  The solid lines show the actual code performance, while the dashed lines show the ideal scenario (i.e. a scaling that decreases as $t \sim 1/N_{\rm proc}$ where $t$ is the wall clock time, and $N_{\rm proc}$ is the number of processors.  These tests were performed on a single 32-core node.   The  combination of the fixed overhead costs and the highly scalable Monte Carlo radiative transfer are apparent in Figure~\ref{figure:scaling}: as the problems become increasingly difficult, and spend more relative time in the radiative transfer, their scaling approaches the ideal limit.

\subsection{Code Comparisons}

In this section, we present code comparisons between \pd \ and
\skirt. {\sc skirt} \citep{baes11a,baes15a,camps15a,verstocken17a,camps20a} is
a state-of-the-art open-source code designed to perform continuum
radiative transfer in dusty systems that has been widely used.  For
these tests, we design {\sc skirt} models to mimic as closely as
possible {\sc powderday} models for a code comparison.  This said, we caution
though that due to differences in the stellar population synthesis
models, dust models, and grid construction, a true apples-to-apples
comparison is currently intractable given the design of both codes.

The galaxy models used for this comparison are {\sc GizmoDisk}, {\sc
  GasolineDisk}, {\sc ChangaMW} and {\sc tnghalo}.  Note, for the {\sc
  arepo} comparison, the {\sc skirt} model was run with the dust
geometry distributed in an octree mesh due to technical difficulties
with {\sc skirt}, while the \pd \ simulation was run over a
reconstructed Voronoi grid tessellated about the gas points.

For the \pd \ models for all, we simulated
the stellar spectrum with \fsps \ with a \citet{chabrier03a} initial
mass function and Padova isochrones
\citep{bertelli94a,girardi00a,marigo08a}.  The dust is distributed
with a constant dust to metals ratio of $0.25$ with \citet{draine03a}
opacities ($R_{\rm V} = 3.1$).  The contribution of PAHs and nebular
lines are turned off.

At the same time, our \skirt \ simulation for the code comparison test
sets up a box around the exact same region of particles for each
model, though with stellar SEDs deriving from \citet{bruzual03a}
population synthesis models.  These too are set up with an assumed
\citet{chabrier03a} IMF and Padova isochrones.  Because these are
pretabulated as lookup tables, these assumptions remain fixed and
represent an intrinsic difference in our comparisons..

In Figure~\ref{figure:powderday_skirt_comparison}, we show a
comparison between the \pd \ and \skirt \ SEDs for these models.  The
galaxies are all set at a distance of $1$ Mpc for this comparison.
While there are some subtle differences that owe to the
differing intrinsic stellar population models, as well as dust grid
construction, by and large the comparison between the two codes
reveals consistent results across the modeled wavelength range.  The
large degree of correspondence between \pd \ and \skirt \ for these
model tests is encouraging.

\subsection{Citing the Code}
Fundamentally as a package, \pd \ simply wraps \fsps
\ \citep{conroy10a,conroy10b}, \hyperion \ \citep{robitaille11a}, \yt
\ \citep{turk11a}, and {\sc astropy} \citep{robitaille13a}.  We
therefore request that any users of the code cite these codes and
papers first and foremost, before citing this paper.

\section{Applications}
In what follows, we demonstrate a number of examples and science
applications of \pd. 
   Each of these could be a scientific investigation unto itself. We
   take a relatively shallow approach to each topic, deferring more
   thorough investigations to future work.  We assumed the Padova
   stellar isochrones \citep{bertelli94a,girardi00a,marigo08a} and
   MILES spectral models \citep{sanchezblazquez06a} for each of these
   applications.
\begin{figure}
  \includegraphics[scale=0.5]{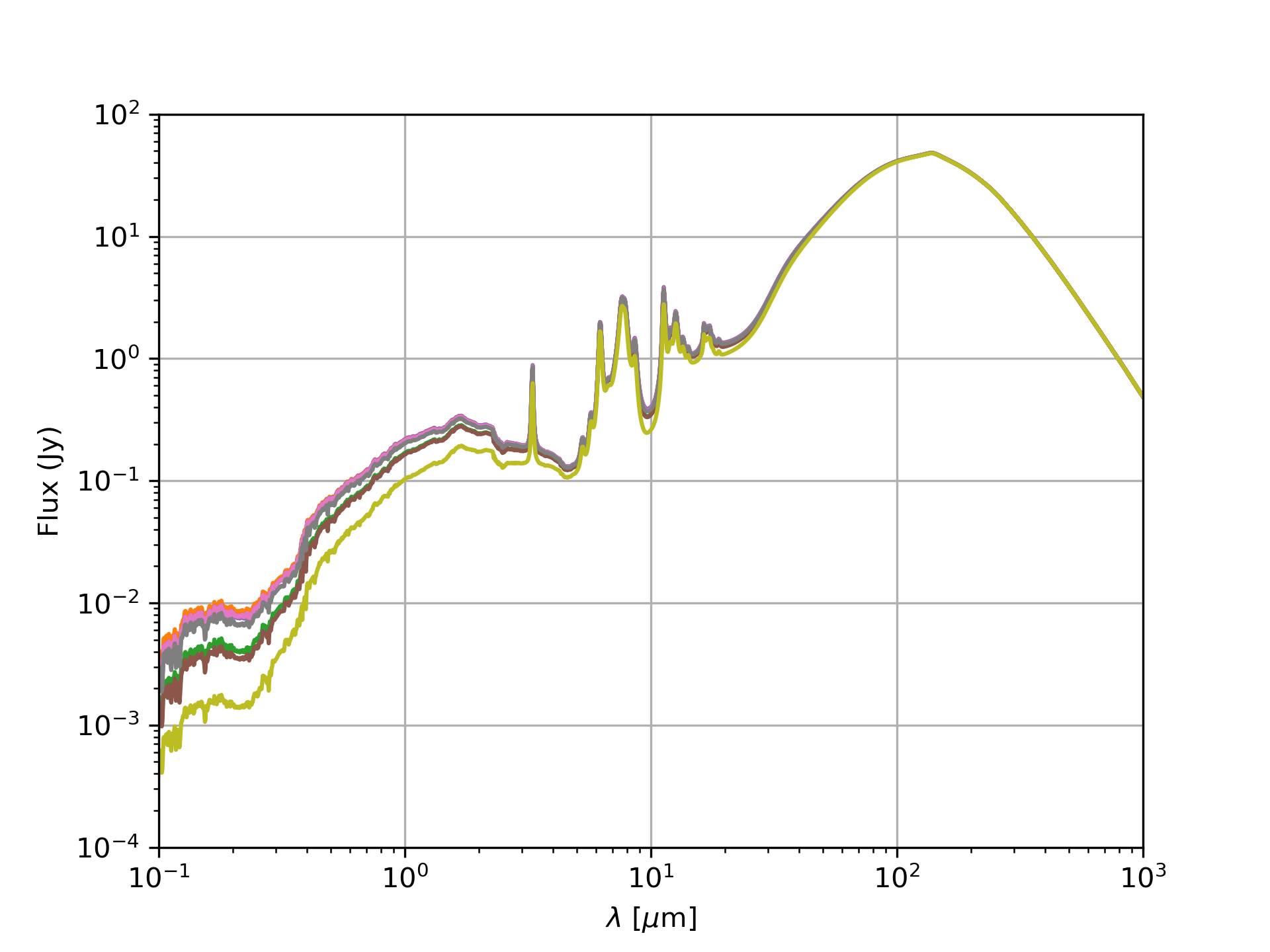}
  \caption{Example panchromatic SED from cosmological zoom-in Milky Way like galaxy {\sc latte} \citep{wetzel16a}.  The different lines denote different viewing angles for the galaxy set at $30$ Mpc. \label{figure:latte_sed}}
\end{figure}

\begin{figure*}
  \includegraphics{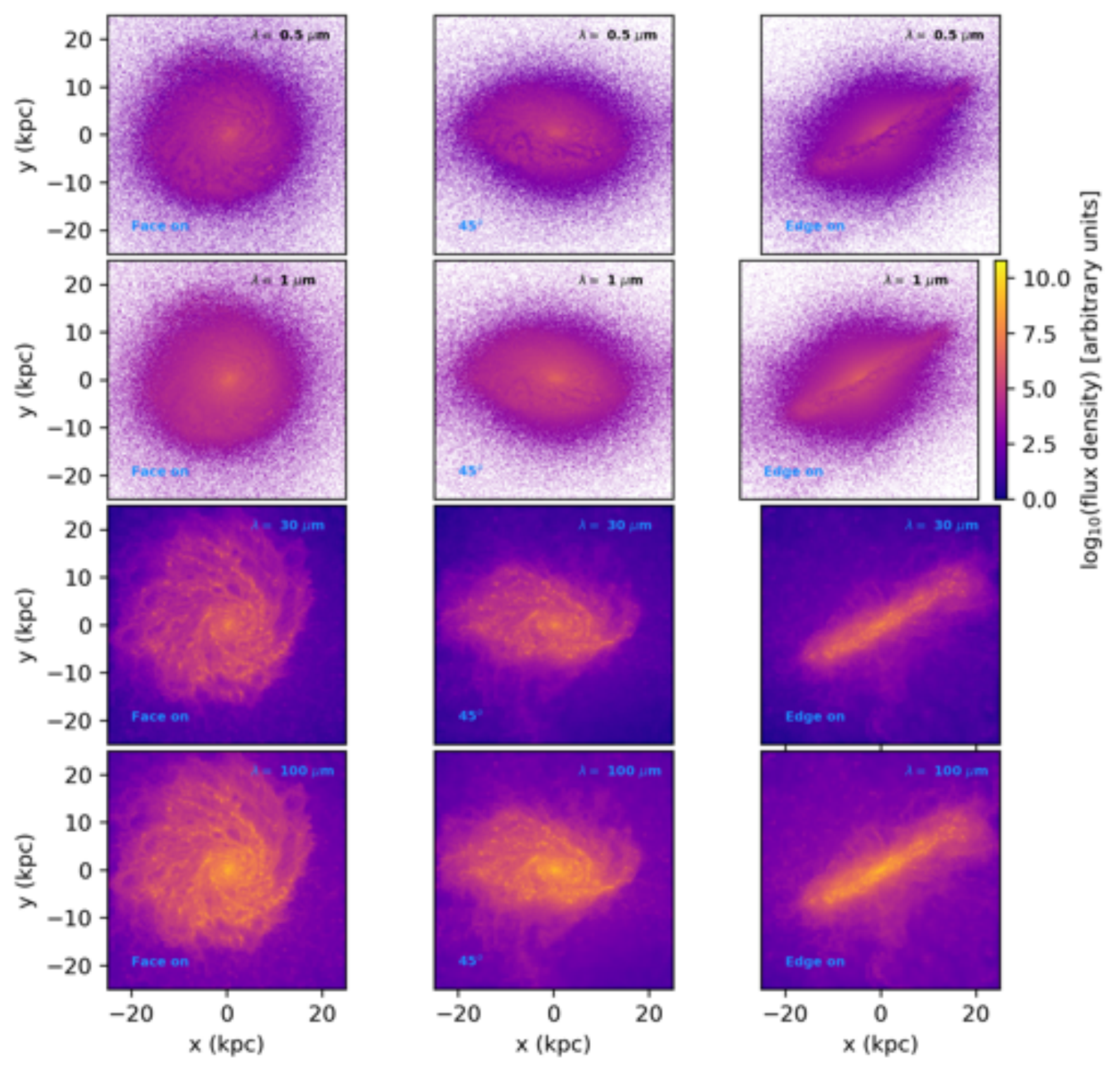}
  \caption{Example monochromatic images of model {\sc latte} at $\lambda =
    [0.5,1,30,100] \mu$m (rows) and at $3$ different 
    viewing angles (columns).  As in Figure~\ref{figure:latte_sed}, the model galaxy is set at $30$ Mpc.  High resolution versions of this image available upon request. \label{figure:latte_image}}
\end{figure*}

\subsection{SEDs and Images}
As an example of the output of \pd, in Figure~\ref{figure:latte_sed}
and Figure~\ref{figure:latte_image}, we show the model SED and
multi-wavelength images for a range of inclination angles for model
{\sc latte}.  The galaxy model is a zoom-in of a Milky Way like galaxy
\citep{wetzel16a}, and the radiative transfer is performed assuming a
\citet{kroupa02a} IMF,  a dust to metals
mass ratio of $40\%$, and PAH emission turned on.

\begin{figure*}
  \includegraphics[scale=0.5]{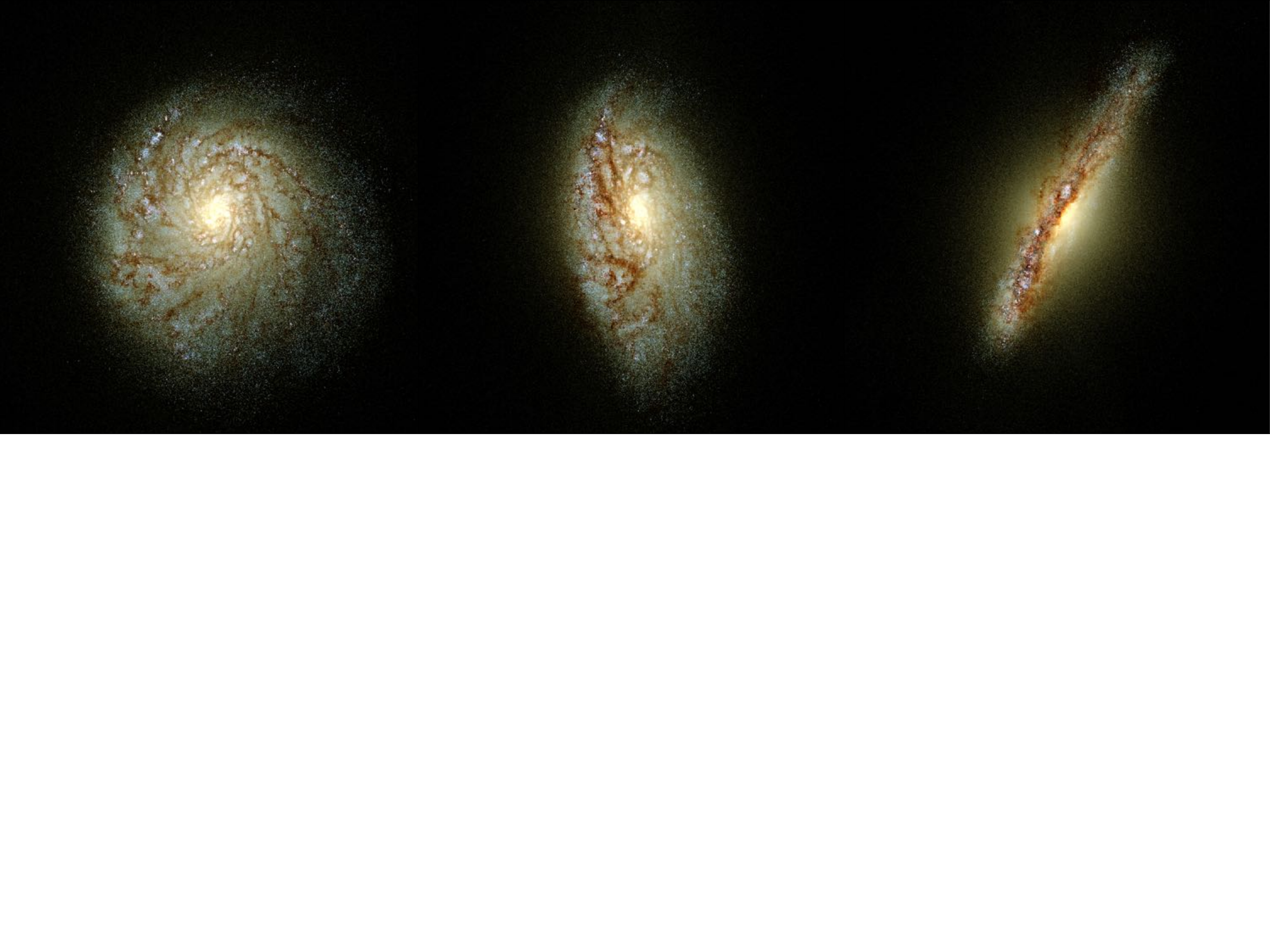}
  \vspace{-7cm}
  \caption{RGB images corresponding to $0.3,0.5$ and $1$ $\mu$m colors from model {\sc latte} for $3$ different viewing angles.  Dust lanes in the galaxy (here, the dust mass is tied to the metal mass as a model assumption) are visible.
  As in Figure~\ref{figure:latte_sed}, the model galaxy is set at $30$ Mpc.  Images generated following the \citet{lupton04a} scalings.  High resolution versions of this image available upon request. \label{figure:snyder_rgb}}
\end{figure*}

The SEDs and images are generated over $9$ viewing angles.  In
Figure~\ref{figure:latte_sed}, we show the viewing-angle dependence of
the SEDs, with more edge-on views naturally resulting in reduced
UV/optical flux. The images are generated at the monochromatic
wavelengths $\lambda=[0.5,1,30,100] \mu$m, and is shown at $3$
different viewing angles.   The images are set at a fiducial
distance of $30$ Mpc.   In the optical/NIR, the face-on views
  highlight the stellar emission, though with clear dust lanes in the
  spiral arms that become more prominent as the angles shift toward
  edge-on.  These dust lanes become significantly more prominent in Figure~\ref{figure:snyder_rgb}, where we show the simulated RGB colors of the same model galaxy (corresponding to $0.3$, $0.5$ and $1$ $\mu$m).  We follow the pipeline of \citet{snyder15a} in generating these images, which uses the scaling techniques described in \citet{lupton04a}.

Generally, \pd \ can generate images at any number of arbitrary
wavelengths.  While the images in Figure~\ref{figure:latte_image} are
not filter convolved, \pd \ ships with a number of canned transmission
filters, and can generate filter-convolved images.

\subsection{Infrared Star Formation Rate Tracers}

The bolometric infrared luminosity is often used as a tracer of the
total star formation rate of a galaxy, with the physical motivation
that the ultraviolet radiation from young newly formed stars is likely
to be absorbed and reprocessed by cold dust in the galaxy.  Often-used
literature calibrations \citep[e.g.][]{kennicutt98a,murphy11a} are
typically developed using population synthesis models with a
relatively simplistic star formation history (or, a stellar population
modeled as a simple stellar population) alongside an assumed dust
covering fraction of the stellar population.  Complications to these
sorts of calibrations include the contribution of older stellar
populations to diffuse dust heating (and therefore adding additional
infrared luminosity that does not originate from young, newly formed
stars), as well as an AGN source term \citep{younger09a,hopkins10a,hayward14a,narayanan15a}.

\begin{figure*}
  \includegraphics[scale=0.55]{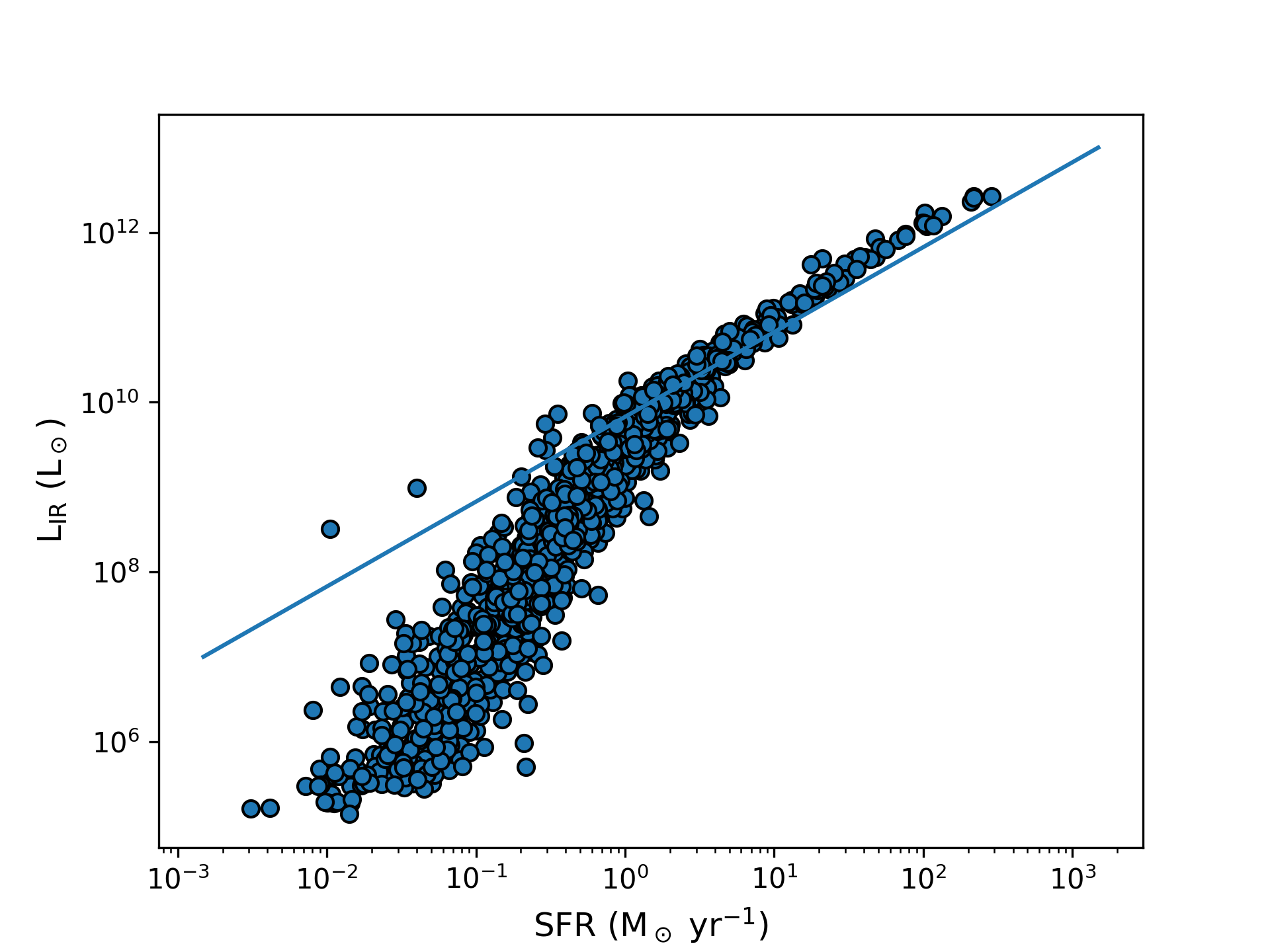}
  \includegraphics[scale=0.55]{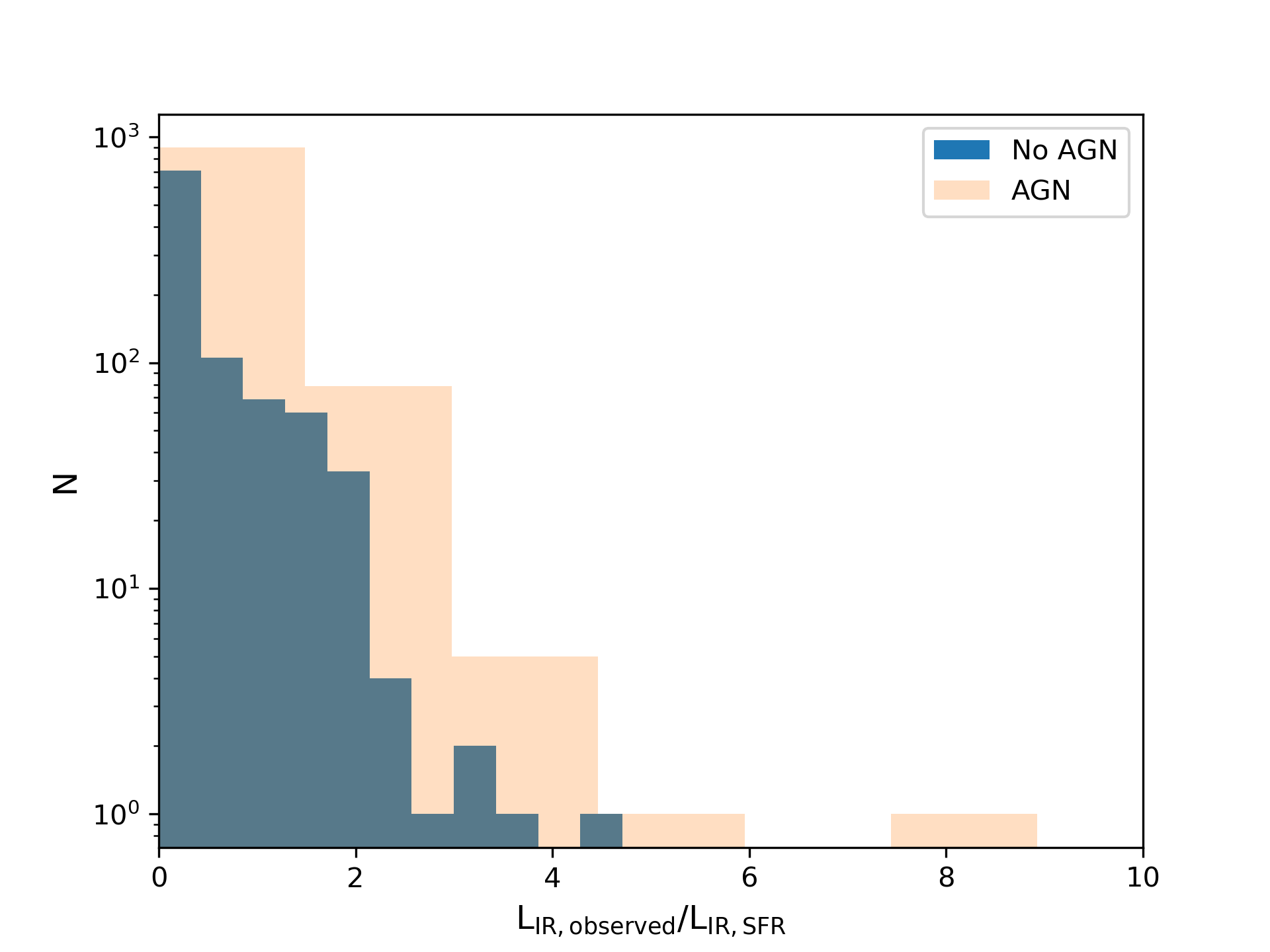}
\caption{{\bf Left:} Infrared luminosity (integrated between $8-1000
    \mu$m) versus star formation rate for the $1000$ most massive
    galaxies at redshift $z=2$ in the {\sc simba m25n512} cosmological
    simulation.  The solid line shows the \citet{murphy11a}
    SFR-L$_{\rm IR}$ relation as compiled in \citet{kennicutt12a}.
    Below SFRs $\la 1$ M$_\odot$ yr$^{-1}$, a lack of dust in galaxies
    drives a precipitous drop in the infrared lumionsity with respect
    to SFR.  {\bf Right:} The impact of AGN on the $L_{\rm IR}$-SFR
    relationship in galaxies.  Histograms show the ratio of the
    synthetic ``observed'' infrared luminosity to what one would
    expect from the \citet{murphy11a} relationship, given the model
    galaxy's SFR.  As is clear, the AGN drives excess power toward
    high luminosities, which may result in over estimates of the SFR
    using a canonical L$_{\rm IR}$-SFR relation for some
    galaxies. \label{figure:sfr_lir}}
\end{figure*}

As an example of the potential of {\sc powderday} in investigating
this issue, we compute the SEDs from the $1000$ most massive galaxies
in the redshift $z=2$ snapshot of the {\sc simba m25n512} cosmological
galaxy formation simulation.  In the left panel of
Figure~\ref{figure:sfr_lir}, we compare the integrated infrared
luminosity for this sample of galaxies ($8-1000 \mu$m in the galaxy's
rest frame) to its $50$ Myr averaged star formation rate.  The solid
blue line shows the \citet{murphy11a} calibration between $L_{\rm IR}$
and the SFR \citep[as reported by][]{kennicutt12a}.

At sufficiently low star formation rates, the modeled galaxies diverge
from the \citet{murphy11a} relation due to a lack of dust.  This lack
of dust can arise both in relatively low mass galaxies, as well as
massive quenched galaxies that have reduced dust masses per unit
stellar mass owing to thermal sputtering \citep[e.g.][]{li19a}.  At
SFRs $\ga 1 M_\odot$ yr$^{-1}$, however, the dust content rises
sufficiently that the bolometric infrared luminosity serves as a
reasonable tracer of the SFR based on empirical calibrations.  At the
highest luminosities modeled here ($\sim 10^{11} L_\odot$), there is some tendency for the infrared
luminosity to overestimate the SFR by a factor $\sim 2$, owing to the
contribution of old stars.  This effect is relatively small given the
limited maximum SFRs in this small {\sc simba} box.
\citet{narayanan15a} demonstrated via cosmological zoom simulations of
massive galaxies that factors of $\sim 3-4$ increase in the infrared
luminosity from old stars can be possible in the most extreme
(i.e. SFR $\sim 1000 M_\odot$ yr$^{-1}$) cases.

In the right panel of Figure~\ref{figure:sfr_lir}, we examine the
impact of including an AGN as a radiating source.  For this example,
we assume that the black hole SED follows a \citet{hopkins07a}
spectrum, and that the luminosity of the black hole is $L = \eta
\dot{M} c^2$, where the efficiency $\eta = 0.1$.  The blue shaded
region shows a histogram of the observed infrared lumionsity compared
to what is expected given the galaxy SFR and the \citet{murphy11a}
SFR-$L_{\rm IR}$ calibration for the same $1000$ galaxies as in the
left panel of Figure~\ref{figure:sfr_lir}.  The salmon shaded region
shows the same histogram for the same galaxies, but including AGN.  As
is clear, there is significantly more power toward large infrared
luminosities, reflecting the impact of the AGN on the increased
$L_{\rm IR}$, and potential for overestimate in inferred SFR.

\subsection{Circumstellar AGB Dust in Low-Metallicity Galaxies}

\begin{figure*}
  \includegraphics[scale=0.55]{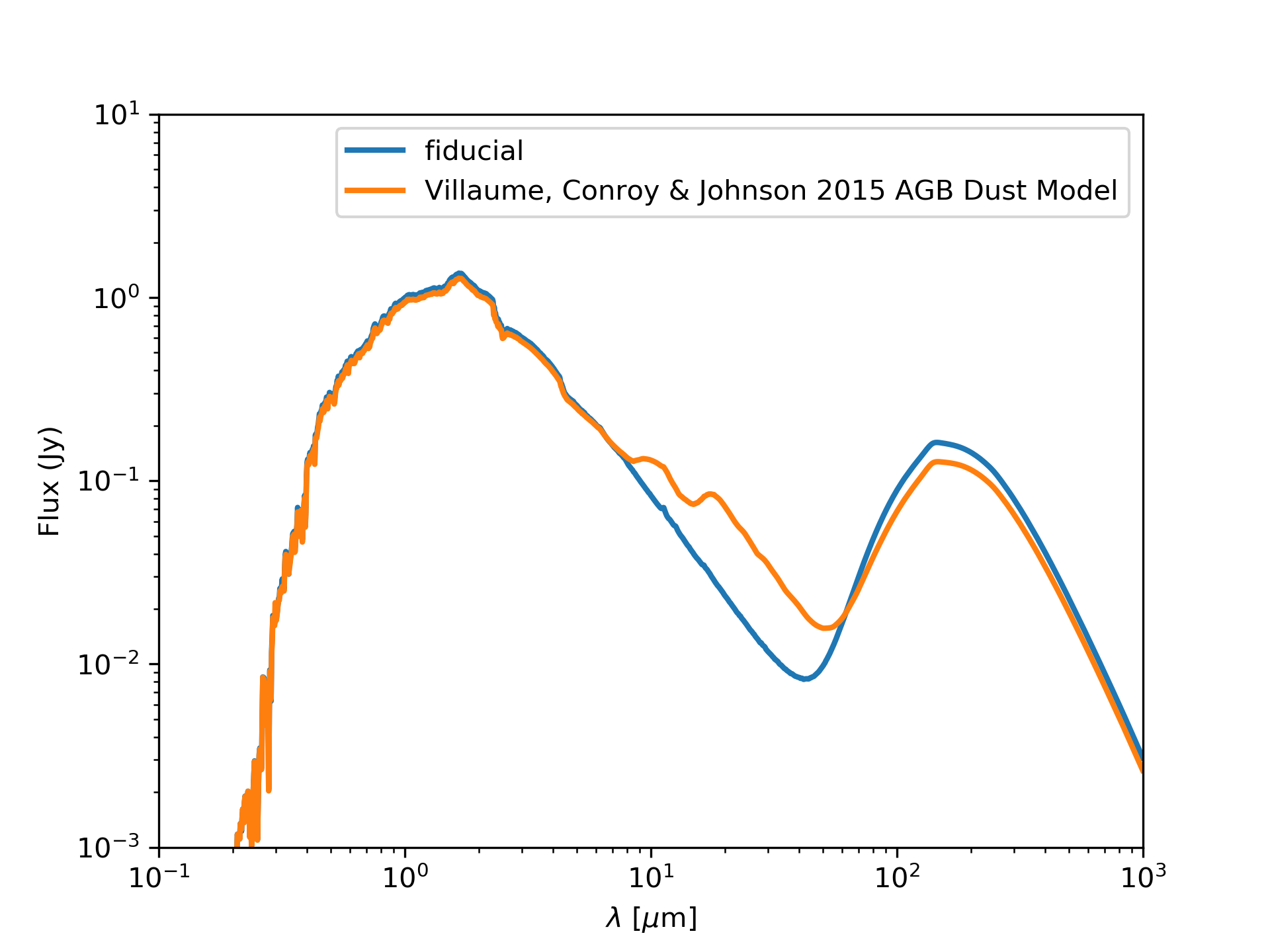}
  \includegraphics[scale=0.55]{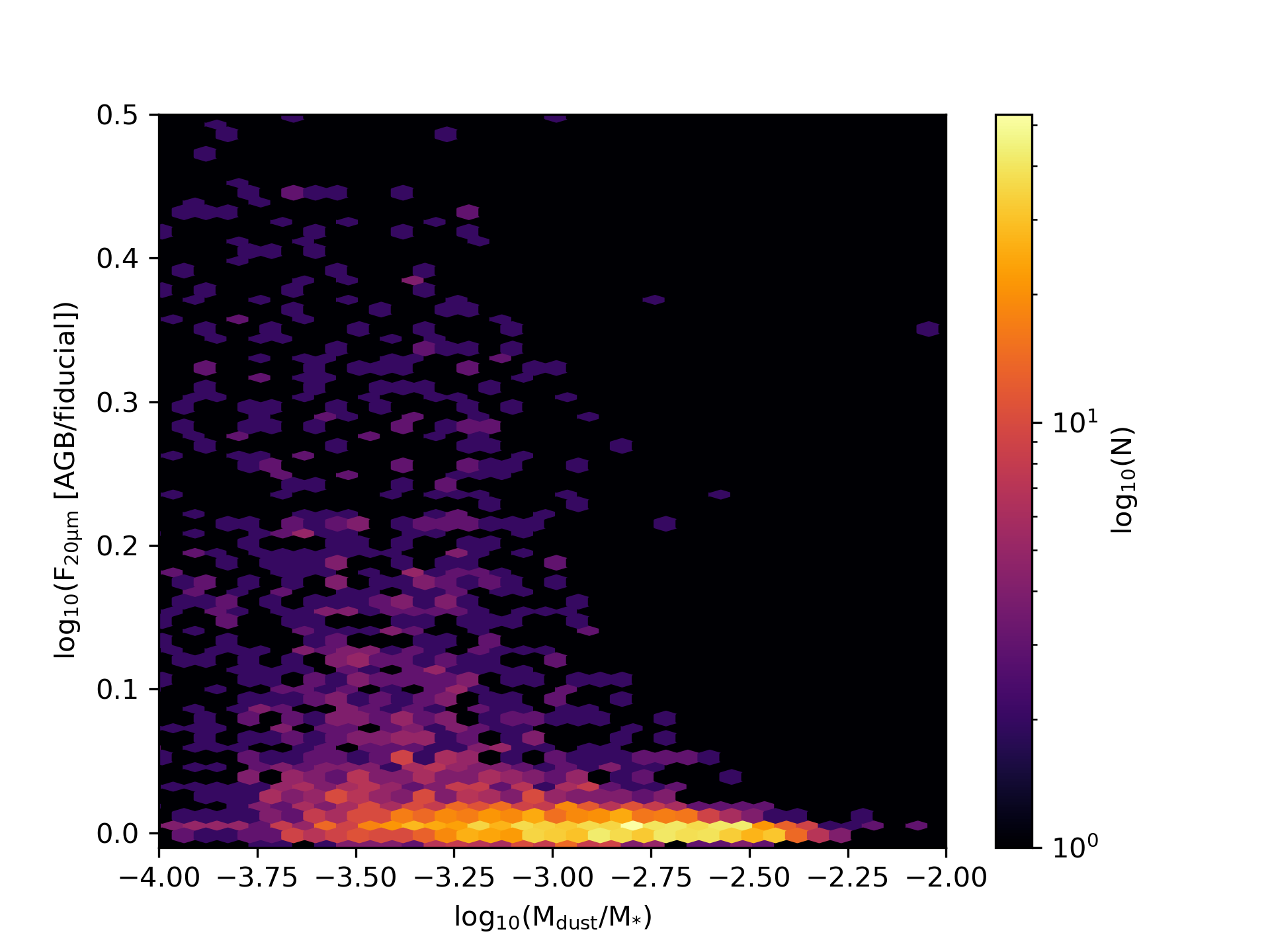}
  \caption{Impact of circumstellar AGB dust on the mid-IR SEDs of
    galaxies as a demonstration of the impact of stellar physics on
    {\sc powderday} radiative transfer models.  {\bf Left:} Example
    SEDs (without PAH models) of a quiescent galaxy with a low dust
    mass fraction from the {\sc simba m25n512} simulation both
    including (orange) and not including (blue) the
    \citet{villaume15a} model for circumstellar AGB dust.  As is
    evident, in this galaxy with relatively low diffuse dust, the
    mid-IR flux is dominated by circumstellar AGB dust.  {\bf Right:}
    Generalization of the left plot for $3000$ galaxies at $z=[0,1,2]$
    from the same {\sc simba m25n512} model, where we show the ratio
    of the $24 \mu$m flux density (as a proxy for the mid-IR) for a
    model with and without AGB dust vs. the fractional dust content of
    the galaxy.  Colored points show a heat map of individual galaxy
    snapshots, with color bar on the right. When the diffuse dust
    content is sufficiently large, the flux in the mid-IR is dominated
    by diffuse dust, and the ratio of mid-IR flux for both models
    converges to $1$.  \label{figure:agb}
  }
\end{figure*}

One of the more powerful aspects of {\sc powderday} is its generation
of stellar SEDs on the fly (as opposed to via lookup tables).  While
this represents a fairly significant computational expense compared to
utilizing lookup tables for the stellar SEDs, the trade-off is
flexibility in being able to explore the impact of stellar physics on
the emergent SED from galaxies.

To demonstrate an example of this, in Figure~\ref{figure:agb}, we show
the impact of including one such aspect of the underlying stellar
model: the effect of circumstellar dust around AGB stars.
\citet{villaume15a} developed {\sc dusty} radiative transfer models
\citep{ivezic97a,ivezic99a} for AGB-phase stars that couple directly
with the {\sc fsps} population synthesis code.  As a result, this is
trivially implementable in the {\sc powderday} framework.  In the left
panel of Figure~\ref{figure:agb}, we show the SED for an arbitrarily
selected galaxy (galaxy 1) with a relatively low $M_{\rm dust}/M_*$
ratio from the {\sc simba m25n512} cosmological simulation.
As is clear, AGB dust can have a significant impact on the mid-IR SEDs
of galaxies with a relatively low dust content.

As has been discussed by previous works
\citep[e.g.][]{silva98a,villaume15a}, the impact of AGB dust on
the mid-IR SED is dependent on the amount of diffuse dust in a galaxy.
As the diffuse dust content increases, this takes over the
contribution of circumstellar dust surrounding AGB stars in the
mid-IR.  To show this, we have modeled the SEDs from the top 1000
  most massive galaxies from the redshifts $z=[0,1,2]$ snapshots from
the {\sc simba m25n512} cosmological simulation both with and without
the contribution of circumstellar AGB dust.  To simplify analysis, we
have not included our model for PAH emission
(c.f. \S~\ref{section:pahs}).  In the right panel of
Figure~\ref{figure:agb}, we plot the ratio of the $24 \mu$m (to serve
as an arbitrary mid-IR wavelength) flux density for a model with AGB
dust turned on to a model without AGB dust turned on as a function of
the fractional dust mass in the galaxy.  At low fractional dust
content ($M_{\rm dust}/M_* \ll 1$) the fractional contribution of the
circumstellar dust dominates in the mid-IR.  As the diffuse dust
content of the galaxy increases, however, the relative contribution of
AGB dust decreases, and the ratio of the $24 \mu$m flux density for
galaxies modeled with and without AGB dust converges to unity.

\subsection{Inclination-Dependent Dust Attenuation Laws}

The attenuation curve of a galaxy represents the effective amount of light lost from a source (typically, understood to mean stellar light), and is a fundamental quantity of interest in SED fitting \citep[see][for a recent review of attenuation laws in galaxies]{salim20a}.  The attenuation curve reflects both extinction along the line of sight (due to absorption and scattering), as well as the contributions by both light scattered back into the line of sight and unobscured stars.  It is therefore highly dependent on the intricacies of the star-dust geometry in a particular system, as well as the viewing angle.  The dust attenuation law on galaxy-wide scales has been the subject of a number of theoretical investigations in recent years, in large part due to advances in coupling hydrodynamic galaxy simulations with dust radiative transfer packages \citep[including \pd][]{jonsson06a,seon16a,narayanan18b,trayford20a}

\begin{figure*}
  \includegraphics[scale=0.5]{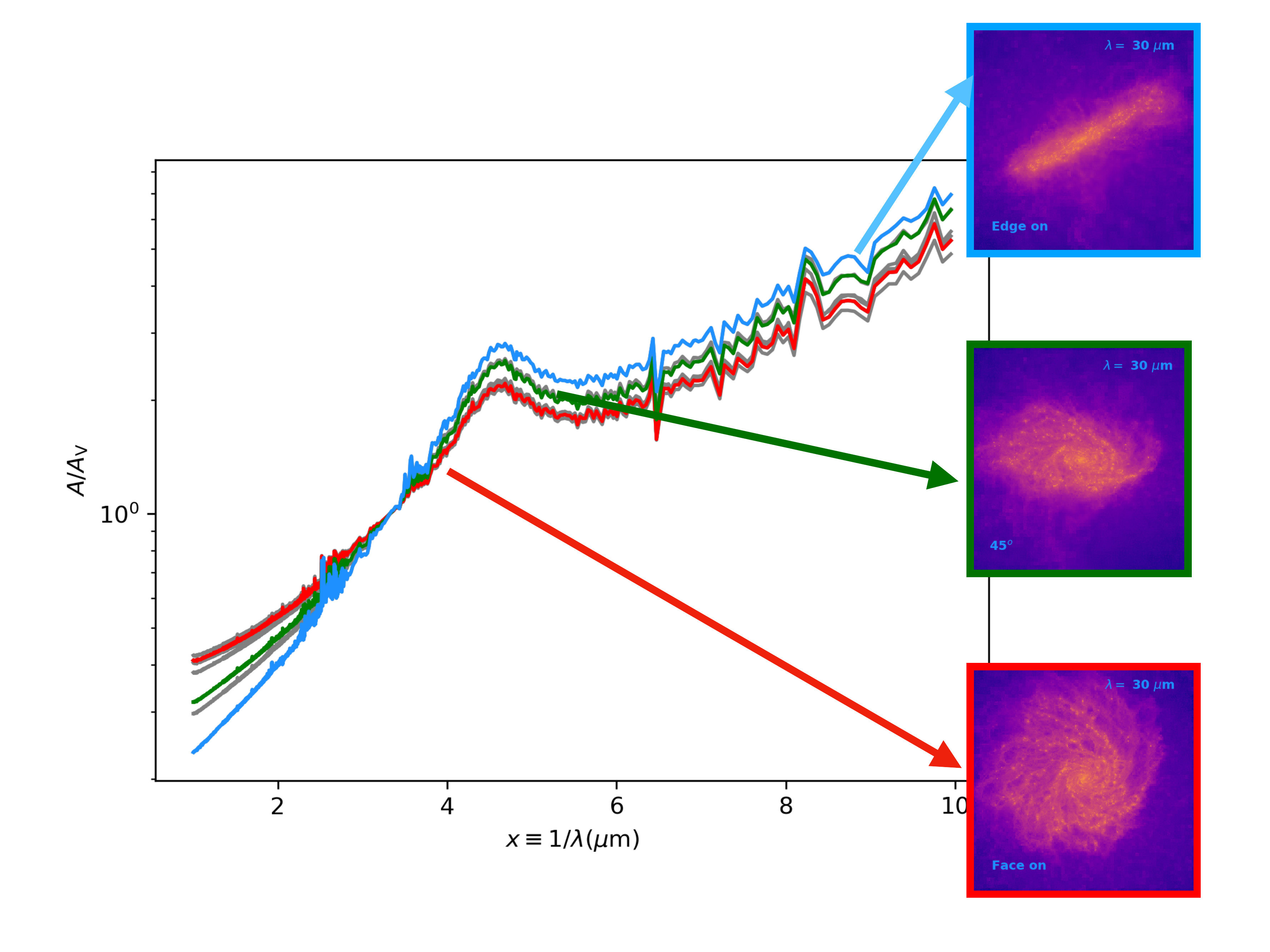}
  \caption{Impact of galaxy inclination angle on attenuation laws.  Curves represent UV-NIR attenuation curves for model galaxy {\sc latte} for $9$ viewing angles, with $3$ particular angles highlighted in color that correspond to the images on the right (face-on, edge-on and intermediary). Edge on inclination angles display steeper normalized curves, while face-on views result in grayer (flatter) attenuation curves. \label{figure:attenuation}}
\end{figure*}

In Figure~\ref{figure:attenuation}, we investigate the role of galaxy inclination on the attenuation law.  For this, we employ the {\sc latte} simulation, as it has a clear disk-like morphology at its final redshift of $z=0$, thus facilitating analysis.   We calculate the attenuation law over $9$ isotropic viewing angles.  While we plot all viewing angles (in light grey), we highlight three particular angles in color that correspond to face-on, edge-on, and an intermediate angle, and show their corresponding $30\mu$m images.   The attenuation curves are normalized at $3000$\AA{} to highlight changes in the slope of the law.

As is clear, there are a diverse range of ultraviolet slopes for the differing viewing angles of model galaxy {\sc latte}.  As discussed in detail in the \citet{salim20a} review, these sorts of slope variations are typically ascribed to variations in the star-dust geometry between galaxies.   Here, we demonstrate that slope changes in the attenuation curve can also be ascribed to the viewing angle of the galaxy: more edge on views have larger differential attenuation in the UV than face-on views.  This point is amplified by the recent observational surveys of \citet{salim18a} and \citet{battisti17b}, and literature references therein. 

\section{Discussion}
\subsection{Is 3D Dust Radiative Transfer Really Necessary?}
We now ask the simple question: is dust radiative transfer actually
necessary to capture the intricacies of the SEDs from galaxies?  To
quantify this, we compare the results
of {\sc powderday} dust radiative transfer modeling of galaxies
(necessarily in $3$ dimensions) to simplified screen models, akin to
what is traditionally used in SED fitting, or generating mock SEDs of
galaxies from population synthesis modeling.

To develop this comparison, we have run {\sc powderday} on the top
$1000$ most massive galaxies in the $z=0$ snapshot of the {\sc simba
  m25n512} cosmological simulation.  We then extract the star
formation histories and metal enrichment histories at relatively high time
resolution ($\sim 150$ evenly spaced time bins over a Hubble time) of
these galaxies, and use those as inputs in {\sc fsps} to build a
composite stellar population.  Using the methods built into {\sc
  fsps}, we then place this composite stellar population behind a dust
screen with \citet{draine07a} dust properties.  For this model, we fix
the variable parameters in the \citet{draine07a} model to their
defaults ($U_{\rm min} = 0.1$, $q_{\rm PAH} = 3.5$ and $\gamma =
0.01$), noting that some modern SED fitting software packages have the
ability to vary these parameters when modeling infrared SEDs
\citep[e.g][]{leja17a,leja19a,johnson19a,johnson19b}.

\begin{figure}
  \includegraphics[scale=0.55]{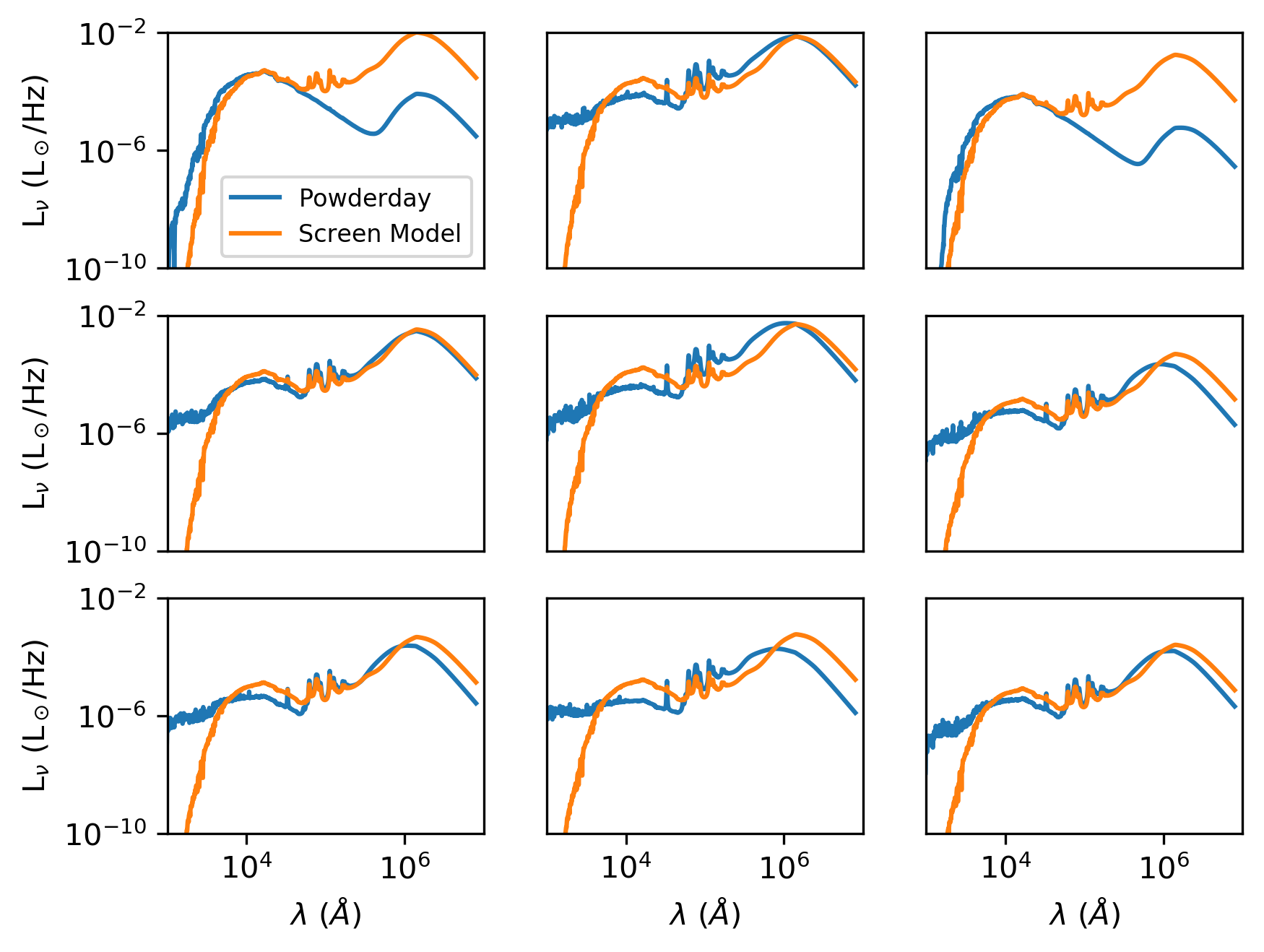}
  \caption{Comparison of 3D dust radiative transfer models of $9$ randomly selected galaxies in the {\sc simba m25n512}
    cosmological simulation (at $z=0$) [blue] compared against 1D dust
    screen models [orange].  By and large, screen models do
    not capture the complexities of the star-dust ISM as 3D models do,
    and therefore over attenuate the UV and optical
    radiation.  \label{figure:fsps_pd_multi}}
\end{figure}

\begin{figure}
  \includegraphics[scale=0.55]{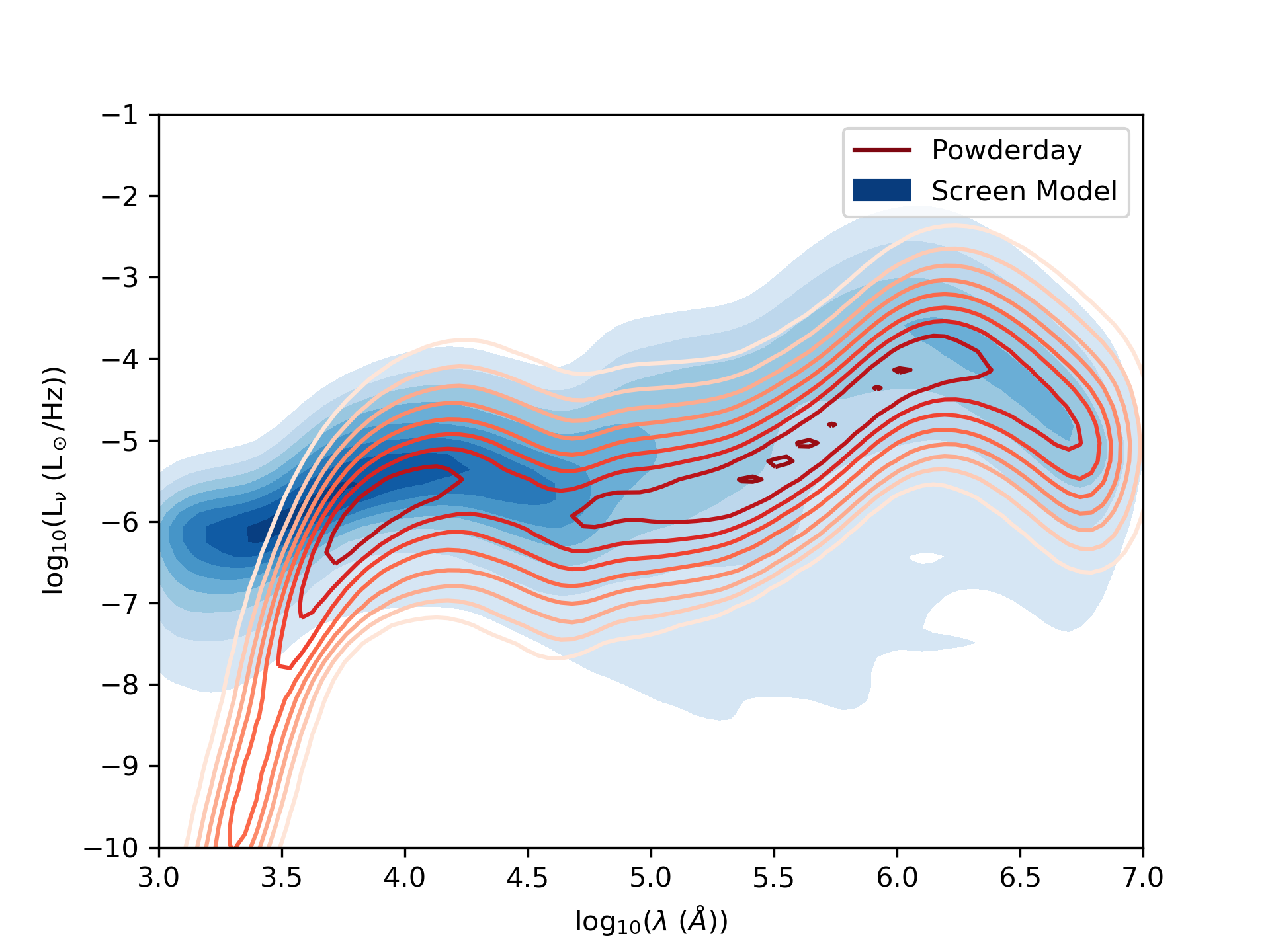}
  \caption{Same as Figure~\ref{figure:fsps_pd_multi}, though showing the results from the top $1000$ most massive galaxies in the {\sc simba m25n512} cosmological simulation at $z=0$.  The blue are the {\sc powderday} $3$D dust radiative transfer models compared against 1D dust
    screen models [red].  The colors/contours show the density of the number
    of models with a given SED shape.   \label{figure:fsps_pd}}
\end{figure}

We show the results of these mock SED models in
Figures~\ref{figure:fsps_pd_multi}, which  shows the SEDs of a random selection of $9$ galaxies.  In Figure~\ref{figure:fsps_pd}, we show the results from all of the modeled galaxies as a density plot.   There are fairly substantial difference in the modeled UV-optical
SED.  The traditional screen model effectively treats all of the stars
in the galaxy as a single source, and all of the dust as a single site
of obscuration.  By neglecting the complex mixing of gas and dust in
galaxies, and the consequent impact on the effective optical depths
\citep[e.g.][]{narayanan18a,narayanan18b}, the simplified screen model
tends to over-attenuate the UV and optical regime of the SED.
Conversely, the 3D dust radiative transfer modeling exhibits
significantly more leakage in the UV, an effect that can 
impact the dust attenuation curve \citep[][]{salim20a},
as well as manifestations in related relations such as the IRX-$\beta$
relation in galaxies \citep{popping17a}.

\subsection{Future Directions}

While {\sc powderday} is an extremely flexible dust radiative transfer
package that contains a number of state-of-the-art algorithms (thanks
to the continued development of the software that it bundles), there
are a number of future code development directions that would be
valuable.

A natural direction  forward would be to include  models for molecular
and atomic line emission from  neutral and molecular gas.  Simulations
with software  such as  {\sc despotic} \citep{krumholz13a}  that model
emission  from  photodissociation  regions   and  molecular  gas  have
demonstrated that  line emission from these  regions typically depends
on  (alongside the  physical properties  of the  gas itself,  which is
typically returned  from a  given hydrodynamic galaxy  simulation) the
cosmic   ray    flux   and    the   incident   UV    radiation   field
\citep[e.g.][]{narayanan17a,li18a}.  Generally, simulations of line emission
from galaxy  simulations have had  to employ local  approximations for
the latter quantity \citep[][see also \citet{olsen18a} for a review of
  some of  these issues.]{olsen15a,olsen17a,leung20a}, neglecting the  impact of
diffuse dust  on large  scales.  This quantity  can, in  principle, be
derived explicitly from dust  radiative transfer modeling.  A valuable
addition  would  therefore  be  to  incorporate  this  information  in
modeling the atomic and molecular  line emission from galaxy formation
simulations.

In the future, we additionally envision {\sc powderday} taking
advantage of modern algorithm developments in the modeling of dust in
galaxy formation simulations.  As we discussed in
\S~\ref{section:dust_models}, {\sc powderday} can already include models
for spatially varying dust content that derives directly from
hydrodynamic simulations \citep*[e.g.][]{mckinnon16a,li19a}.  However,
newer approaches are now allowing for the evolution of a size
distribution of grains that vary spatially across the galaxy
\citep[][Qi Li et al., in
  prep.]{hirashita15a,mckinnon18a,aoyama18a,hou19a}.  When coupling
these size distributions with an assumed extinction efficiency, these
spatially-varying size distributions can be used to model the dust
extinction law.  A major step forward for dust radiative transfer
codes will be to allow for the ability to include non-uniform
extinction laws as returned from models such as these, in order to
develop ever-more realistic SED models from galaxies.

\section{Summary}
We have presented the first release of the \pd \ dust radiative
transfer package, which is designed to extract synthetic broadband
SEDs and nebular line emission from hydrodynamic idealized and
cosmological galaxy formation simulations.  \pd \ is designed with
ease-of-use and flexibility in mind, aiming to eliminate interface
hurdles between practitioners of different hydrodynamic galaxy
formation methods and radiative transfer.  In particular, some major
features of \pd \ include:
\begin{enumerate}
\item \pd \ is designed with a high level of flexibility in mind.
  Users have the ability to vary many aspects of the stellar emission
  via the bindings to the flexible \fsps \ population synthesis code,
  nebular line emission from \ion{H}{2} regions, and models for the dust
  content in galaxies.
  \item We have implemented front ends in \pd \ to seamlessly
    interface with a number of hydrodynamic galaxy formation packages
    (through the use of \yt \ as an intermediary), including {\sc
      gizmo}, {\sc arepo}, {\sc gasoline}, {\sc changa} and {\sc
      enzo}.      
    \item By leveraging \hyperion \ as our dust radiative transfer
      solver, we are able to maintain a high level of scalability, and perform the radiative transfer
      over octree, adaptive and voronoi meshes.
      \item We include multiple models for optional emission from AGN.
\end{enumerate}
We have additionally demonstrated the capabilities of \pd \ via a number of 
scientific applications:
\begin{enumerate}
\item We have examined the relationship between the SFR and infrared
  luminosity ($8-1000 \mu$m) in a sample of cosmologically simulated
  galaxies.  While the bulk of the galaxies fall on the
  \citep{murphy11a} relationship between $L_{\rm IR}$ and SFR,
  galaxies at very low SFRs deviate significantly owing to a lack of
  dust.  At the high SFR end, the implied SFR from the $L_{\rm IR}$
  can exceed the true SFR due to both a contribution of older stars to
  the heating of diffuse dust, as well as (when present) contribution
  from AGN (Figure~\ref{figure:sfr_lir}).

  \item We have modeled the contribution of circumstellar AGB dust to
    the mid-infrared flux of galaxies using the \citet{villaume15a}
    model for AGB dust emission.  For quiescent galaxies with a
    relatively low dust content, circumstellar AGB emission can
    provide a significant boost to the mid-IR flux.  As the $M_{\rm
      dust}/M_*$ ratio increases, however, the impact of AGB dust emission becomes negligible (Figure~\ref{figure:agb}). 

  \item We investigated the role of inclination angle on the integrated dust attenuation law from a model disk galaxy.  Edge on views of the galaxy tend to show steeper attenuation laws in the ultraviolet, while face-on views result in grayer (flatter) curves.

    \item We have compared the results from $3$D dust radiative
      transfer to simplified screen models
      (Figure~\ref{figure:fsps_pd}).  Generally, full $3$D radiative
      transfer shows more power at UV and optical wavelengths (and
      consequently, less in the infrared) due to a more complex
      star-dust geometry than screen models typically allow for.
      
\end{enumerate}

\section*{Acknowledgements}In the course of developing \pd, numerous
individuals have provided substantial help.  First and foremost, we
are eternally grateful to the hundreds of members of the \yt
\ community both for their contributions to the \yt \ codebase, as
well as for their participation and positive-natured support in the
\yt \ email lists and Slack page.  Without this inclusive community,
\pd \ would have never been possible. We thank Nell Byler, Romeel
Dav\'e, Ross Fadely, Kevin Flaherty, Nathan Goldbaum, Fabio Governato,
Chris Hayward, Cameron Hummels, Ben Keller, Dusan Keres, Patrick
Sheehan, Rachel Somerville, Brian Svoboda, and Greg Walth for helpful
conversations along the way.  We are grateful to Oscar Agertz, Mike
Butler, Ji-hoon Kim, Keita Todoroki, Kentaro Nagamine and James
Wadsley for their willingness to share sample \agora \ snapshots in
advance of their publication (which enabled early {\sc powderday}
testing), and Mike Tremmel, Nicole Sanchez, Andrew Pontzen, Paul
Torrey, and Federico Marinacci for sharing {\sc changa} and {\sc arepo
  smuggle} sample outputs for testing.  We thank the countless
scientists who have beta tested \pd \ over the years, including:
Hollis Akins, David Ball, Derrick Carr, Charlotte Christensen, Joel
Christian, Jarren Jennings, Katarina Kraljic, Reilly Millburn, Emily
Moser, Justin Otter, Gergo Popping, Spencer Scott, Emery Trott and
David Zegeye. D.N. thanks Vicente Rodriguez-Gomez, Paul Torrey,
Federico Marinacci, and Laura Blecha for their assistance in working
with {\sc arepo} \ simulation outputs.  We thank the {\sc fire-2}
collaboration for their having made the {\sc latte} simulation results
public.  D.N. additionally expresses deep appreciation toward Joe
Cammisa at Haverford College, and the UF Research Computing group for
their roles in maintaining the Fock and HiPerGator high performance
computing clusters, respectively, where the main development of {\sc
  powderday} took place. Partial support for DN was provided by NSF
grants AST-1009452, AST-1442650, and NASA HST AR-13906.001 from the
Space Telescope Science Institute, which is operated by the
Association of Universities for Research in Astronomy, Incorporated,
under NASA contract NAS5-26555, and a Cottrell College Science Award
awarded by the Research Corporation for Science Advancement. AJK
acknowledges an STFC studentship grant ST/S505365/1.  JHW is supported
by NSF grants AST-1614333 and OAC-1835213 and NASA grants NNX17AG23G
and 80NSSC20K0520. MJT was supported in part by the Gordon and Betty
Moore Foundation's Data-Driven Discovery Initiative through Grant
GBMF4561.  Finally, D.N. expresses deep appreciation for the Aspen
Center for Physics, at which the idea for (and naming of, on a
bluebird day at Ajax Mountain) {\sc powderday} came about.



\bibliographystyle{mnras}
\bibliography{./pd_refs}


\end{document}